\documentclass{aa}  
\usepackage{booktabs} 
\usepackage[colorlinks=true,linkcolor=blue,citecolor=blue,filecolor=blue, urlcolor=blue]{hyperref}
\usepackage{graphicx}
\usepackage{txfonts}
\begin{document}

   \title{How the presence of a giant planet affects the outcome of terrestrial planet formation simulations}

   \titlerunning{Terrestrial planet formation with outer giants}

   \author{Zhihui Kong,
          \inst{1}\inst{2}
          Anders Johansen,
          \inst{2}
          Michiel Lambrechts,
          \inst{2}
          Jonathan H. Jiang
          \inst{3}
          \and
          Zong-Hong Zhu
          \inst{1}
          }

    \authorrunning{Kong, Z.H., Johansen, A. et al.}
   \institute{Department of Astronomy, Beijing Normal University,
              Beijing, People's Republic of China\\
              \email{201831160004@mail.bnu.edu.cn}
         \and
             Center for Star and Planet Formation, Globe Institute, University of Copenhagen, ØsterVoldgade 5-7, 1350 Copenhagen, Denmark\\
             \email{}
        \and
            Jet Propulsion Laboratory, California Institute of Technology, Pasadena, USA
             \email{Jonathan.H.Jiang@jpl.nasa.gov}
             }

   \date{Received; accepted}


    \abstract{The architecture and masses of planetary systems in the habitable zone could be strongly influenced by outer giant planets, if present. We investigate here the impact of outer giants on terrestrial planet formation, under the assumption that the final assembly of the planetary system is set by a giant impact phase. Utilizing a state-of-the-art N-body simulation software, GENGA, we interpret how the late stage of terrestrial planet formation results in diversity within planetary systems. We design two global model setups: in the first we place a gas giant on the outer side of planetesimals and embryos disk, while the other only has planetesimals and embryos but no giant. For the model including the outer giant, we study the effect of different giant initial masses, in the range 1.0-3.0 Jupiter mass, and orbital radii, in the range 2.0-5.8 AU. We also study the influence of different initial positions of planetesimals and embryos on the results. Our N-body simulation time is approximately 50 Myr. The results show that the existence of outer giant will promote the interaction between planetesimals and embryos, making the orbits of the formed terrestrial planets more compact, but placing the giant planet too close to the planetesimals and embryos disk suppresses the formation of massive rocky planets. In addition, under the classical theory, where planetary embryos and planetesimals collide to form terrestrial planets, our results show that the presence of a giant planet actually decreases the gap complexity of the inner planetary system.}


   \maketitle
%

\section{Introduction}

    Over the past three decades exoplanet researchers have discovered 
    more than five thousand planets outside our Solar System \citep{Winn_2015}.
    We have seen a huge diversity in the exoplanet population, which is 
    reflected in many types of planets discovered, such as super-Earths, hot gas giants, 
    cold gas giants, ice giants and terrestrial planets. The orbits and eccentricities of planets in multiple systems provide another window into exoplanetary systems \citep{Lissauer_2011,Weiss_2018}. Therefore, this motivates us to generalize terrestrial planet formation models to account for the diversity of exoplanet systems.
 
    According to the standard model, planet formation can be divided into three stages: growth from dust to pebbles by coagulation, assembly of planetesimals from pebbles and growth to embryos and fully-formed planets by mutual impacts, pebble accretion and gas accretion \citep{Raymond_2004,Raymond_2006,Levison_2010,Johansen_2015, Levison_2015, Mordasini_2015, wurm_2018, Johansen_2021}. The focus of this article lies on the stage of terrestrial planets formation by impacts, specifically addressing the process by which planetary embryos grow into terrestrial planets. In the classical model, terrestrial planets form through to form through a series of violent collisions among numerous planetary embryos and planetesimals \citep{Wetherill_1990,Chambers_1999}. Several studies show that this process requires approximately 30-100 Myr for Earth to form \citep{Chambers_2001,John_2010, Raymond_2014}. Nevertheless, when pebble accretion is taken into account, the Earth can form much faster within the protoplanetary disc in $3$$\sim$$5$\,Myr \citep{Liu_2018, Lambrechts_2019, Broz_2021, Johansen_2021}. Gas giants must form rapidly within the protoplanetary disc under all circumstances. However, analyses of numerous star clusters of different ages indicate that the typical timescale for the protoplanetary disk is only a few Myr, which is earlier than the inner terrestrial planet formation stage \citep{Chambers_1998,Haisch_2001,Brice_2001,Hillenbrand_2008,Hasegawa_2012,Nesvorn_2021}. Therefore, the presence of giant planets must significantly influence the process of terrestrial planet formation. In this paper, we explore the formation of terrestrial planets by giant impacts after the dissipation of the protoplanetary disc. We critically compare two models: one incorporates an external gas giant in the formation of terrestrial planets, while the other excludes any giants. This way we investigate the role of gas giant planets in the shaping the masses and orbits of inner terrestrial planet systems.

    N-body simulation methods have proven to be highly valuable for understanding the orbital patterns of Earth-like planets within our Solar System, such as the differences and similarities between Venus and Earth and between Earth and Mars \citep{Raymond_2004,Hansen_2009, Nesvorn_2021, Woo_2022, Batygin_2023}. Meanwhile, several studies explore not only the planetary architecture within our Solar System but also study the growth of rocky planets during giant planet instabilities, aiming to explain the diversity of exoplanets \citep{Raymond_2011,Raymond_2012,Carrera_2015}. In this paper, we explore the presence of a giant planet as a free parameter in terrestrial planet formation, with a mass range of 1--3 Jupiter mass and an orbital range between 2--5.8 AU. \cite{Nesvorn_2021} investigated the impact of a giant planet instability on the formation of terrestrial planets and found that accelerates the growth of the terrestrial planets. This implies that the instability of giant planets likely occurred during the process of terrestrial planet formation \citep{Clement_2018,Clement_2019}. The authors examine various instability scenarios previously found to align with numerous Solar System constraints. They discovered that resonances with giant planets contribute to the removal of solids available for accretion near 1.5 AU, thereby impeding the growth of Mars. These authors, approaching with a view centered on the Solar System, used their the N-body simulation to match the structure of terrestrial planets in the Solar System. In addition, some work has focused how a migrating chain of giant planets affect the terrestrial planets formation \citep{Fogg_2005, Fogg_2007,Raymond_2006b,Edward_2008}. In our study, however, we conduct a multitude of N-body simulations to explore the structural diversity of exoplanet systems and elucidate the impact of outer gas giants on the variety observed in inner terrestrial planetary systems.
    
    Furthermore CPU-based N-body simulation codes have served as the primary tool for investigating planet formation; however, due to computational constraints, planetesimals in these simulations are usually treated as test particles that can only interact with and feel the gravitational force from more massive embryos, but can not mutually interact with each other \citep{Brien_2006, Walsh_2011, Jacobson_2014, Brasser_2016, Quintana_2016}. This approximation can significantly lower the total number of active particles in the system and keep the computational time reasonable. However, such simplified simulation settings could lead to strong limitations in fully understanding the process of terrestrial planet formation, especially for modelling the growth of Mars. \cite{Woo_2022} used GPU-based algorithms to conduct simulations over extended time periods in systems comprising a substantial number of particles. In our study, we use the $GENGA$ code by \cite{Grimm_2014} to accelerate the simulation process. We carried out a comprehensive analysis using a total of 140 simulation samples, wherein we systematically varied the initial state of the gas giant. Each sample simulation time is about 50 Myr, which basically covers the evolution process from the planetary embryo to the protoplanet.
    
    In the study by \cite{He_2023}, the $Kepler$ Giant Planet Search (KGPS) as detailed by \cite{Weiss_2023} was employed to investigate the influence of outer gas giants on the masses and orbits of interior, smaller planets. The findings indicate that systems with an outer giant planet exhibit a higher level of gap complexity. Following that approach, we applied a comparable methodology to analyze our simulation samples. We do not include here the effects of the gas giant formation phase and its early migration on the velocity dispersion of the planetesimals in the inner regions of the disc, although these are potentially important \citep{Walsh_2011, Guo_2023}. These aspects will be the focus of our future research.
    
    We outline this paper as follows: In Section 2, we introduce the numerical method and describe the inner disk of planetesimals and embryos and the orbital parameters of the outer gas giant. In Section 3, we present the results of the N-body simulations. We investigate the impact of initial states of giant planets on the formation of terrestrial planets and present our findings in Section 3.1. We describe the concept of gap complexity concept to analyze these simulation results in Section 3.2. A statistical analysis is performed on the simulation samples in Section 3.3. In Section 4, we present a comprehensive summary of the simulation results for three distinct types of samples representing the final stage of terrestrial planet formation in Section 4.1. We succinctly outline the limitations of our model and provide a forward-looking perspective on potential avenues for future research endeavors in Section 4.2. 
\section{Method}

In this section, describe how we construct a 3D global N-body simulation model based on classical terrestrial planet formation model to investigate the influence of outer giants on the terrestrial planet formation. We focused on the study of collisions between planetesimals and planetary embryos, while disregarding the migration of the giant planet.

\subsection{Introducing the N-body codes}
\begin{figure}
   \centering
   \includegraphics[width=0.5\textwidth]{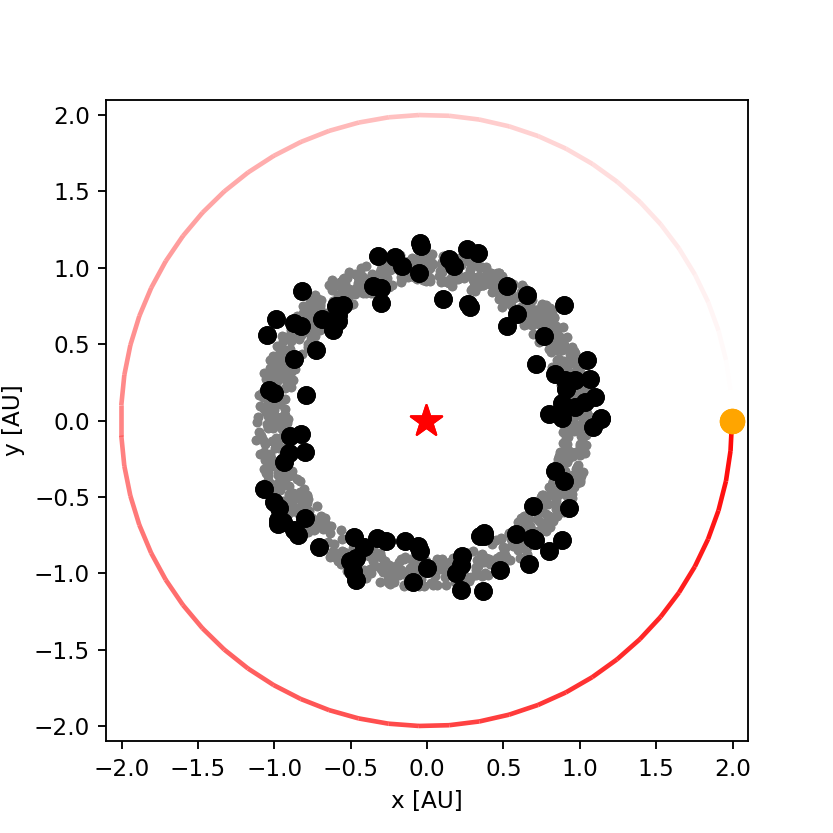}
   \caption{The image shows a simulation setup using the REBOUND code, featuring a central star with a solar mass. Orbiting the star are 100 planetary embryos, each with a mass of 0.05 Earth masses (represented by black dots) and 1000 planetesimals, each with a mass of 0.005 Earth masses (represented by gray dots), arranged in a ring approximately 1 AU in radius. Additionally, a gas giant (represented by an orange dot) orbits at a distance of 2 AU.}
    \label{pl_001}%
    \end{figure}

The N-body simulation software $REBOUND$, extensively employed in the field of astrophysics, contains both a non-symplectic integrator, with adaptive time stepping \citep{Everhart_1985,Rein_2012,Rein_2015} and a hybrid symplectic integrator \citep{Rein_2019}, that is similar to the $MERCURY$ \citep{Chambers_1999}. Consequently, $REBOUND$ has evolved into a pivotal research tool for investigating planet formation and orbital dynamics \citep{Lam_2018, Kong_2022}. It is straightforward to set up a model with this code, such as Fig. \ref{pl_001} that illustrates a planetary system comprising a solar mass, 100 planetary embryos and 1000 planetesimals positioned at approximately 1 AU, and accompanied by a gas giant-sized placed in an exterior orbit. However, the $REBOUND$ code, which utilizes the CPU as the fundamental arithmetic unit, exhibits suboptimal performance in long-term simulations of high particle numbers. This is true despite that it also provides a parallel computing scheme.

Leveraging the GPU as the fundamental computational unit in $GENGA$, the inherent parallel computing logic facilitates enhanced computational speed, rendering it an efficient tool for investigating the last stage of terrestrial planets formation by numerous collisions between planetary embryos and planetesimals. Moreover, $GENGA$ enables simulation over extended duration. However, constructing a model using this code necessitates inputting the position and velocity data for each individual particle, which proves inconvenient when dealing with models comprising a substantial number of particles. Hence, in this study, we initially employ $REBOUND$ to set up a range of planetary systems with diverse initial conditions, evolving the model parameters in the requisite $GENGA$ format for subsequent extensive simulations.

\subsection{Model Setup}

Previous research has shown that a narrow annulus of planetesimals and embryos is conducive to the formation of Earth-mass terrestrial planets \citep{Hansen_2009, Kaib_2015, Raymond_2017, Nesvorn_2021, Izidoro_2021, Joiret_2023}. Therefore, we designed two models, one with a giant planet on the outside of the disk, and one with no giant. We set the central star to 1 solar mass and uniformly place 100 planetary embryos of 0.05 $M_ \oplus$ between 0.8 and 1.2 AU, for a total mass of 5 $M_ \oplus$, and uniformly place 1000 planetesimals of 0.005 $M_ \oplus$ between 0.9 and 1.1 AU for a total additional mass of 5 $M_ \oplus$. The eccentricity of embryos and planetesimal are set according to a Rayleigh distribution with mean value of 0.01, while their inclinations are set as a Rayleigh distribution with a mean value of 0.005. Here the Rayleigh distribution as function with mean sigma is defined as $ \text{Rayleigh}(\sigma) = \sigma \sqrt{-2 \ln(\text{uniform}[0,1)}\, $
\noindent where uniform[0,1) represents a uniformly distributed random variable between 0 and 1 (with 1 excluded). We note that by using a distribution of inclinations, we consider a 3-D model rather than a plane.

Next, we transfer the initial model into the $GENGA$ code for a simulation time of about 50 Myr. In practice, the calculation step is 6 days and the total number of simulation steps is $3\times 10^9$. This yields a total simulation time of approximately 50 Myr and the file recording interval is about 0.1 Myr. To improve the running speed, the storage interval for the first 10 Myr is set at 0.1 Myr, and for the subsequent 40 Myr, it is changed to 1 Myr. The model parameters are summarized in Table \ref{model_1}, while the parameters for the outer gas giants are presented in the subsequent section 2.3. In this study, planetesimals and embryos are mixed and placed in a narrow annulus, with all particles in the model acting as active particles. When two particles come into close proximity, they undergo collision and either merge or experience a scattering event. We run $GENGA$ with its default a collision condition. The resolution in terms of particle number is comparable to the first "high-resolution" simulations of terrestrial planet formation \citep{Raymond_2006, Brien_2006, Morishima_2008}, but lower than many current simulations. However, we established 140 simulated samples to study the influence of outer giants on the formation of inner terrestrial planets. The parameters for giant stars are set in Section 2.3.

   \begin{table}
      \caption[]{Model parameters}
         \label{model_1}
     $$ 
         \begin{array}{p{0.5\linewidth}l}
            \hline
            \noalign{\smallskip}
            Mass of Central Star      &  1\, \mathrm{M_ \odot} \\
            Total mass of Embryos & 5\, \mathrm{M_\oplus}     \\
            Total mass of Planetesimals           & 5\, \mathrm{M_\oplus} \\
            \noalign{\smallskip}
            \hline
            \noalign{\smallskip}
            Number of Embryos     &    100   \\
            Number of Planetesimals    &    1000   \\
            \noalign{\smallskip}
            \hline
            \noalign{\smallskip}
            Orbits of Embryos     &    0.8 - 1.2\, \rm AU   \\
            Orbits of Planetesimals    &    0.9 - 1.1\, \rm AU  \\
            \noalign{\smallskip}
            \hline
            \noalign{\smallskip}
            Calculation step     &    6\, \rm days   \\
            Total time    &    \approx 50\, \rm Myr    \\
            \noalign{\smallskip}
            \hline
         \end{array}
     $$ 
   \end{table}
\subsection{Configuration of gas giants}

We categorize our samples into two groups: the \textbf{No Giant} group represents scenarios where only planetesimals and embryos are present, without the inclusion of a gas giant in the outer system. The initial states of planetesimals and planetary embryos are randomly generated based on their distribution as discussed in section 2.2. In the \textbf{With Giant} group, we investigate the influence of an outer gas giant on the formation of terrestrial planets within a mass range of 1.0 to 3.0 Jupiter masses with intervals of 0.5 $M_{\rm Jup}$, and orbits ranging from 2.0 to 5.8 AU with intervals of 0.2 AU. In addition, the eccentricity of the gas giant is set to 0.05, since the resulting gravitational perturbations can quantitatively affect the outcome of terrestrial planet accumulation \citep{Chambers_2001, Brien_2006}. Moreover, in our model, the giant is influenced by the inner disk, which can lead to changes in eccentricity over time as illustrated in Fig \ref{pl_002}. Much previous work has noted this phenomenon \citep{Chambers_1998,Chambers_2001,Raymond_2004,Raymond_2006,Brasser_2023}, and here we quantify it in Fig. \ref{pl_012}. Meanwhile, we maintain the same initial state for planetesimal and planetary embryos in each sample. We established an additional set of samples, which we name clone\_001-020, to investigate the impact of the randomized initial states of planetesimals and planetary embryos on the outcome of terrestrial planet formation. In the clone\_001-020 samples, we take the same initial state of a gas giant as giant\_047 (with a giant planet mass of 2.0 $M_{\rm Jup}$ orbiting at 3.2 AU) and randomly generate planetesimals and embryos based on the distribution outlined in the section 2.2. Table \ref{model_2} provides a comprehensive summary of the parameters for the two sample groups.

\begin{table*}
\centering
\caption{Naming convention}
\label{model_2}
\begin{tabular}{@{}llll@{}}
\toprule
\textbf{Group name}        & \textbf{Include samples}                     & \textbf{Mass of giant}       & \textbf{Orbits of Giant}\\ \midrule

\textbf{No Giant}       
        & nogiant\_000-020       &               &                    \\
        \hline
\textbf{With Giant}  
        & giant\_001-020 & $G_{\text{m}}=1.0\, M_{\rm Jup}$ 
        & $G_{\text{a}}(\rm 2AU\text{--}5.8AU)$     \\
        
        & giant\_021-040 & $G_{\text{m}}=1.5\, M_{\rm Jup}$             
        & $G_{\text{a}}(\rm 2AU\text{--}5.8AU)$     \\
        
        & giant\_041-060 & $G_{\text{m}}=2.0\, M_{\rm Jup}$              & $G_{\text{a}}(\rm 2AU\text{--}5.8AU)$     \\
        
        & giant\_061-080 & $G_{\text{m}}=2.5\, M_{\rm Jup}$              & $G_{\text{a}}(\rm 2AU\text{--}5.8AU)$     \\
        
        & giant\_081-100 & $G_{\text{m}}=3.0\, M_{\rm Jup}$              & $G_{\text{a}}(\rm 2AU\text{--}5.8AU)$     \\    
        & clone\_001-020 & $G_{\text{m}}=2.0\, M_{\rm Jup}$             & $G_{\text{a}}=3.2\rm AU$                   \\ 
        \bottomrule
\end{tabular}
\end{table*}


%
\section{Results}
In this section, we investigate how the mass and orbit of the giant planet affects the formation of terrestrial planets. Our model allows for the possibility of planetesimals and planetary embryos colliding with the outer gas giant and recording changes to its mass. However, the outer gas giant s more likely to eject planetary embryos from the system due to the gravitational slingshot effect. We present our main findings in section 3.1. Subsequently, we describe the concept of gap complexity, $C$, from \cite{Gilbert_2020} and follow the approach \cite{He_2023} to analyse our simulation results in section 3.2. Finally, we conducted a statistical analysis of the simulation samples, and the results are shown in section 3.3.

\subsection{The effect of gas giants}
   \begin{figure}
   \centering
   \includegraphics[width=0.5\textwidth]{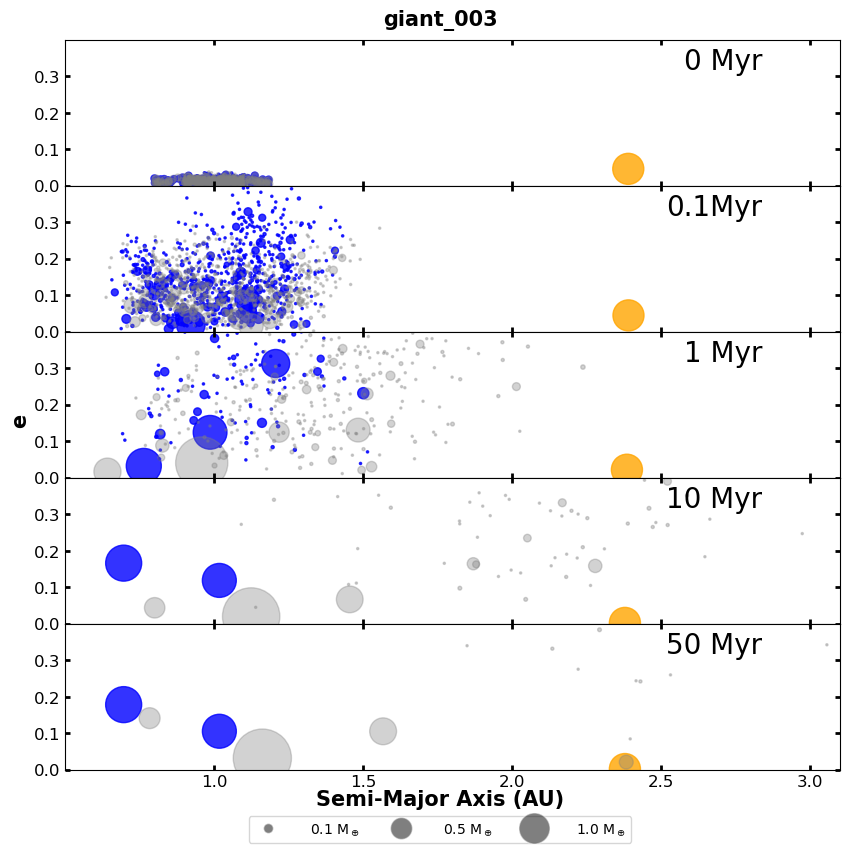}
   \caption{The plot depicts snapshots of terrestrial planet evolution. The orange dot denotes a gas giant with a mass of $1.0\, M_{\rm Jup}$. The blue dots trace the evolution of planetesimals and planetary embryos. Grey dots are used as a background to illustrate the evolutionary paths of terrestrial planets in the absence of a gas giant's perturbation.}
    \label{pl_002}%
    \end{figure}
First, we depict snapshots of giant\_003, where a giant is placed (represented by an orange dot) at 2.4 AU with a mass of $1.0\, M_{\rm Jup}$, in Fig. \ref{pl_002}. We used grey dots as a background to illustrate the evolution of terrestrial planets in the absence of a gas giant's perturbation. The blue dots indicate masses, orbits and eccentricities of planetesimals and planetary embryos. The results show that the eccentricity of the particles increases significantly in the early stages of evolution, which promotes orbital crossing and collisions among the particles. Compared with scenarios lacking an outer giant, the planetary system stabilizes quickly at 10 Myr, with most of the inner mass being ejected from the system by the giant. Additionally, the orbit of the giant is affected by the inner disk, leading to a gradual decrease in its eccentricity. In the end, the two remaining planets, which have a mass similar to Earth's, maintain a high eccentricity because their orbits are close to each other. Under the influence of their gravitational forces, the eccentricity remains between 0.1 and 0.2. We further discuss the eccentricity variations across different models in Section 3.3.
    \begin{figure}
        \centering
        \includegraphics[width=0.5\textwidth]{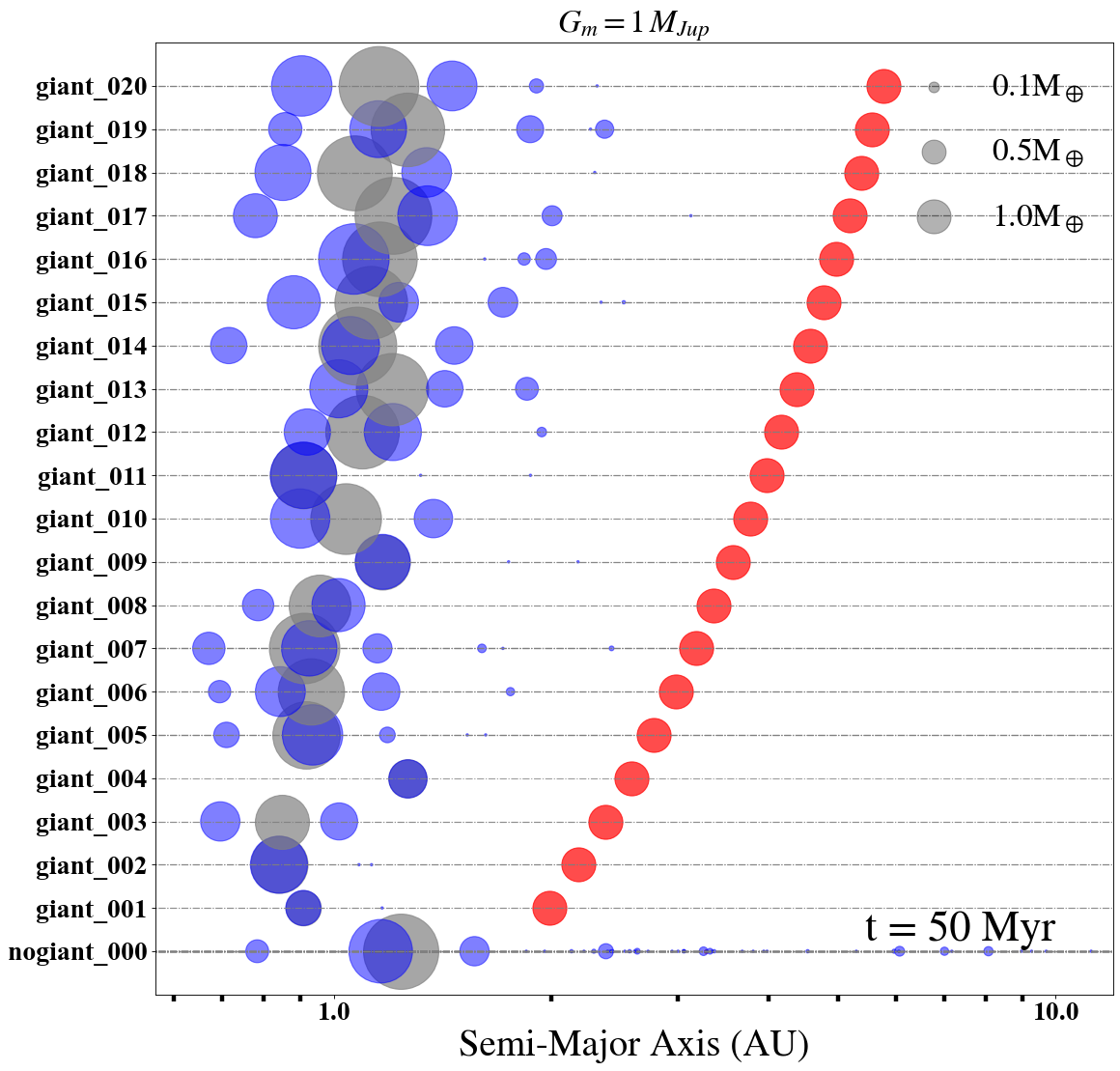}
        \caption{The plot shows the resulting planetary system after approximately 50 Myr, illustrating the effect of placing a Jupiter-mass giant planet (indicated by red dots). The blue dots represent terrestrial planets, with the scale provided in the upper right corner. The position of each red dot indicates the location of the outer gas giant planet. In this plot, the mass of the giant is fixed at $1M_{\rm Jup}$. The position of the gray dot reflects the weighted average radius of the terrestrial planets (blue dots), and the size of the gray dot signifies the total mass of these terrestrial planets. It is important to note that a mass threshold of $\geq \, 0.1 M_ \oplus$, is applied in the calculation of the gray dot.}
        \label{pl_003}
    \end{figure}
    
The results of the samples giant\_001-020 and nogiant\_000 are presented in Fig. \ref{pl_003}. The plot shows the result of the planetary system simulations after a time of approximately 50 Myr. By this stage, only a few massive particles remain, forming the final architecture of the planetary system \citep{Levison_2003, Raymond_2004, Raymond_2006}. The position of the gray dot gives the weighted average radius of the blue dots, while the size of the gray dot represents the sum of the masses of the blue dots. Note that there is a mass cut, $\geq \, 0.1 M_ \oplus$, applied when calculating the gray dot, in order to exclude the influence of any remaining low-mass particles. In addition, there is no giant in the sample nogiant\_000, as shown in the bottom row of Fig. \ref{pl_003}. By comparing nogiant\_000 with giant\_001-020, it becomes evident that the presence of an exterior gas giant leads to a more compact orbital configuration for the inner terrestrial planets. The presence of a giant relatively close to the planetesimal disk suppresses the growth of the terrestrial planets and pushes their location closer to the star. Moving the giant further away from the planetesimal disk, on the other hand, will increase the diversity of terrestrial planets. The gray dots indicate that as the gas giant moves away from the planetesimal disk, there is a gradual increase in both the total mass of the remaining planets within the system and their mean orbital radius.

    \begin{figure}
        \centering
        \includegraphics[width=0.5\textwidth]{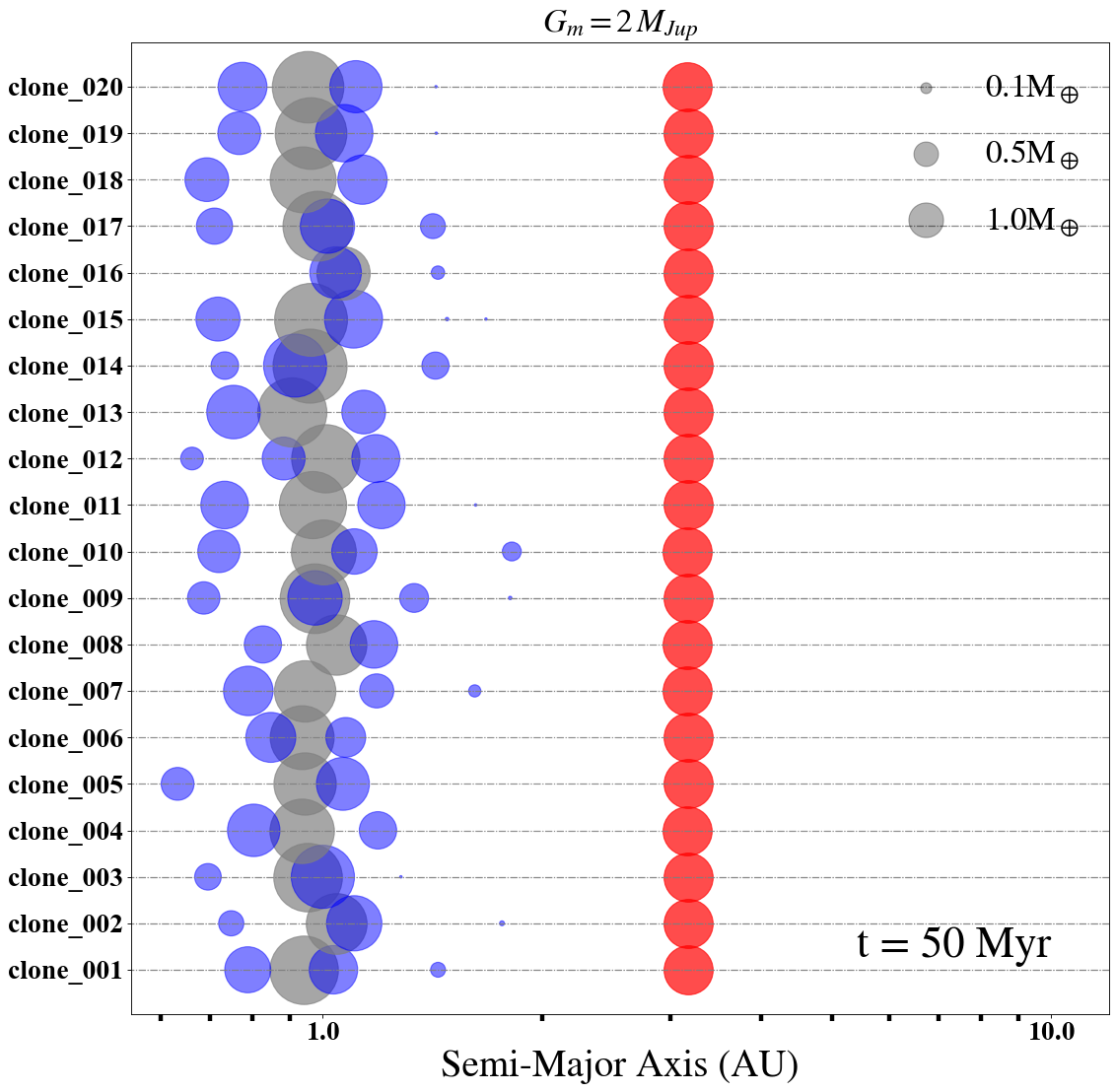}
        \caption{This plot shows the results for the clone\_001-020, which assumes the same initial state of a gas giant and randomly generates planetesimals and embryos based on their distribution, as discussed in Section 2.2. We find that slight differences in the initial state of planetesimals and embryos significantly impact on the resulting planetary systems (indicated by blue dots). However, this influence can be reduced by considering the mean orbit of the planets (represented by gray dots).}
        \label{pl_004}
    \end{figure}

It is necessary to study the statistical properties of N-body simulation, due to the chaotic nature of their evolution \citep{Kokubo_2006}. Consequently, we have set up an additional set of samples, clone\_001-020, to investigate the impact of initial states of planetesimals and planetary embryos on formation outcomes, as depicted in Fig. \ref{pl_004}, while maintaining unchanged conditions for the outer gas giant. The findings demonstrate that random perturbations in the initial state of planetesimals and planetary embryos exert significant influences on the orbits and masses of the resultant terrestrial planets, as evidenced by the distribution pattern depicted by blue dots in the illustration. However, the initial conditions of planetesimals and planetary embryos seem to have little effect on the final total mass and mean orbit of the inner terrestrial planets, as shown in the distribution of gray dots in the figure. Therefore, the total mass and mean orbital characteristics of the inner remnant terrestrial planets in a planetary system can serve as robust physical parameters for investigating the formation process of such planets, thereby mitigating the influence of initial conditions pertaining to planetesimals and planetary embryos on the obtained results.

    \begin{figure}
        \centering
        \includegraphics[width=0.5\textwidth]{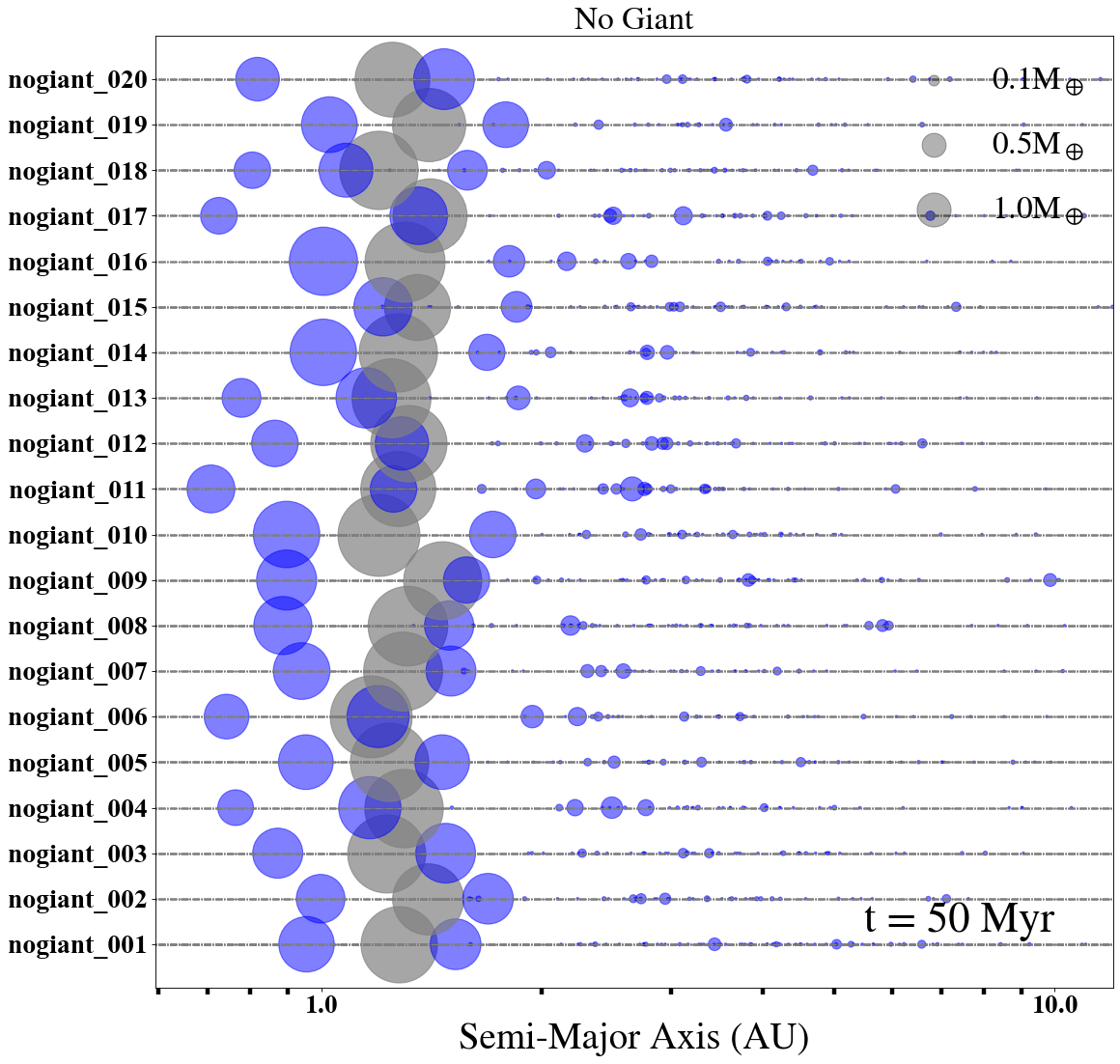}
        \caption{In this plot, we show results in the absence of a giant planet, with the initial states of planetesimals and planetary embryos randomly generated based on their distribution. The findings depicted in the figure are consistent with the points made in Fig. \ref{pl_004}, namely that the total mass and mean position of the formed planets is relatively constant despite the variation in the detailed planetary architectures. Moreover, this plot further underscores that including an outer gas giant leads to a more tightly confined orbit for terrestrial planets formed within its inner boundary, compared with Fig \ref{pl_003}.}
        \label{pl_005}
    \end{figure}

For models that do not contain any external gas giant, we also set up random samples \textbf{No Giant} to study the effects of the initial state of planetesimals and planetary embryos on planet formation, as shown in Fig. \ref{pl_005}. The findings depicted in the figure are consistent with those illustrated in Fig \ref{pl_004}, where random changes in the initial state of planetesimals and planetary embryos have a significant impact on the results. Again, this effect can be significantly reduced by considering instead the total mass and mean orbit of the planets. By comparing Fig. \ref{pl_003} and \ref{pl_005}, it is further emphasized that the inclusion of outer gas giants would result in a more tightly confined orbit for the terrestrial planet formed within its inner boundary.

    \begin{figure}
        \centering
        \includegraphics[width=0.5\textwidth]{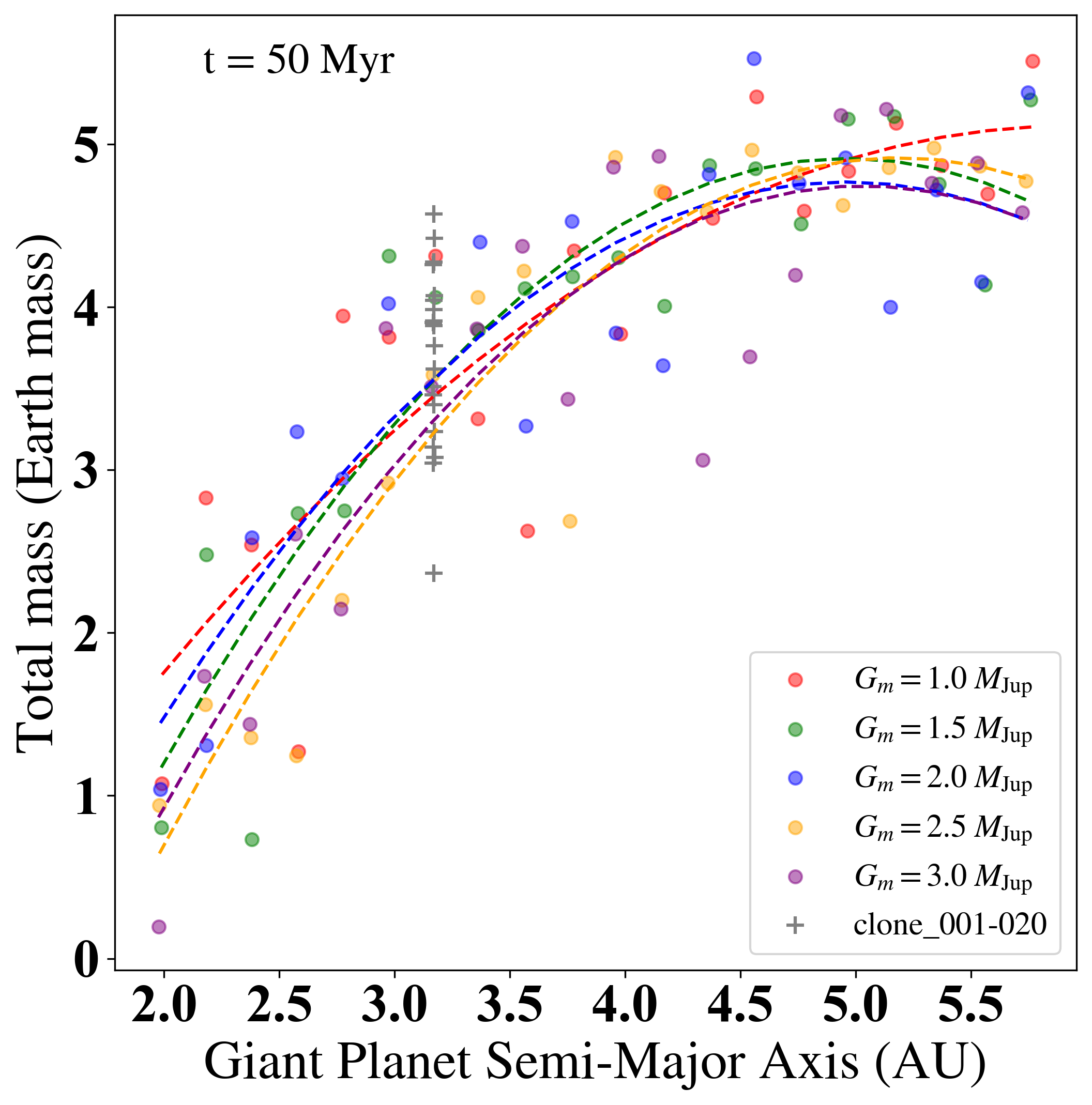}
        \caption{The plot shows the total mass in terrestrial planets at the end of the simulation, as a function of the semi-major axis of the gas giant. The different colored dots represent the effect of different giant planet masses on the simulation results. The plus dots represent the clone\_001-020 with the gas giant at 3.2 AU. There is a clear trend (indicated by a dashed line representing a second order polynomial fit) that the total mass of the remaining terrestrial planets increases as the gas giants move away from the inner planetesimals.}
        \label{pl_006}
    \end{figure}

Next, we concentrate the results of the \textbf{With Giant} group in Fig. \ref{pl_006}. The $x$ coordinate is the position of the initial giant planet, and the $y$ coordinate shows the total mass of all remaining inner terrestrial planets, with $m \geq \, 0.1 M_ \oplus$, in the system. The simulated samples of various gas giant states are represented by dots in different colors. Moreover, polynomials were utilized to individually fit points of the same color, enhancing the clarity of data trends. The next series of plots use the same legend. To enhance the visibility of the outcome, we use dotted lines in corresponding colors to fit the data points in Fig. \ref{pl_006}. There is a clear trend that the total mass of the remaining terrestrial planets increases as the gas giants move away from the inner planetesimals. There is also a weak tendency for the corresponding fitting curve to move downward overall as the mass of the gas giant increases. These phenomena can be attributed to the gravitational interaction between the gas giant and the planetesimal disk. When the gas giant is close to the planetesimal disk, consequently, certain planetesimals and planetary embryos are either captured by or ejected from the system due to their attraction towards the gas giant, resulting in a reduction in mass for the remaining terrestrial planets. As the gas giant moves away from the planetesimal disk, its gravitational influence weakens, leading to an increased mass and orbital diversity among the inner terrestrial planets. Similarly, as the mass of the gas giant increases, its gravitational influence increases and the fitting curve moves downward. Furthermore, the distribution of these samples, clone\_001-020, does not exhibit a distinct segregation from the overall distribution of the samples, giant\_001-100, implying that the initial state of planetesimals and planetary embryos less influence on the mass of the remaining terrestrial planets.

    \begin{figure}
        \centering
        \includegraphics[width=0.5\textwidth]{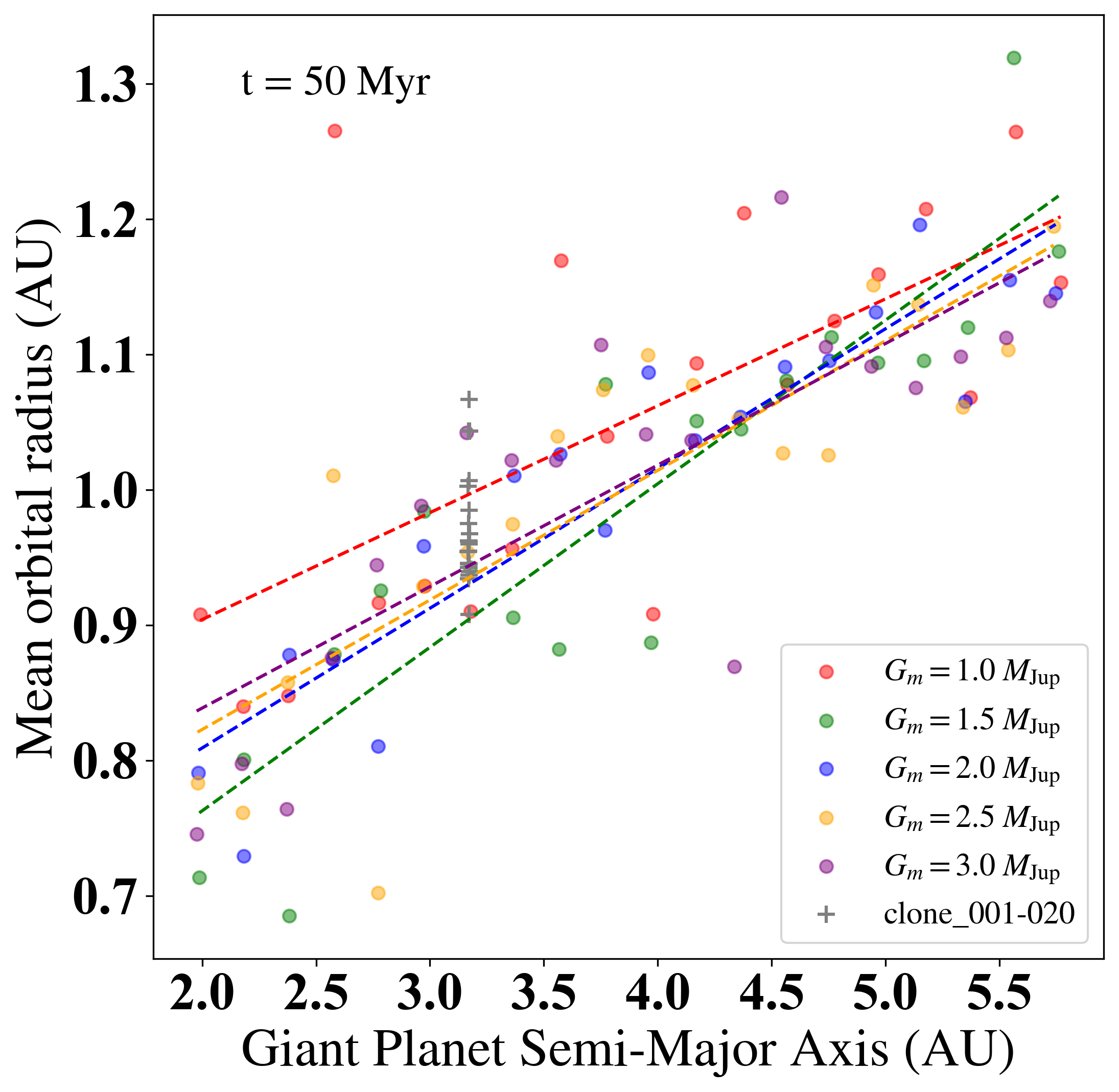}
        \caption{The plot shows the mean orbital radius as a function of the semi-major axis of the giant planets. There is a clear trend (indicated by a dashed line representing a linear fit) that the weighted average of the orbits of the inner terrestrial planets increases as the gas giant moves away from the inner planetesimals.}
        \label{pl_007}
    \end{figure}

In Fig. \ref{pl_007}, we replace the y-coordinate with the weighted average of the orbits of the inner terrestrial planets, defined as $m \geq \, 0.1 M_ \oplus$. There is a clear trend that the weighted average of the orbits of the inner terrestrial planets increases as the gas giant moves away from the inner planetesimals. Meanwhile, the fitted curve drifts downward as the mass of the initial gas giant increases. This also implies that as gas giants decrease in mass or move further away from planetesimals, their gravitational influence on the planetary disk weakens, resulting in an expanded orbital distribution of the formed terrestrial planets. The distribution of the samples , clone\_001-020, exhibits greater concentration in Fig. \ref{pl_007} compared to Fig. \ref{pl_006}, implying that the initial states of planetesimals and planetary embryos exert a diminished influence on the average orbital characteristics of terrestrial planets when compared to the total mass.

    \begin{figure}
        \centering
        \includegraphics[width=0.5\textwidth]{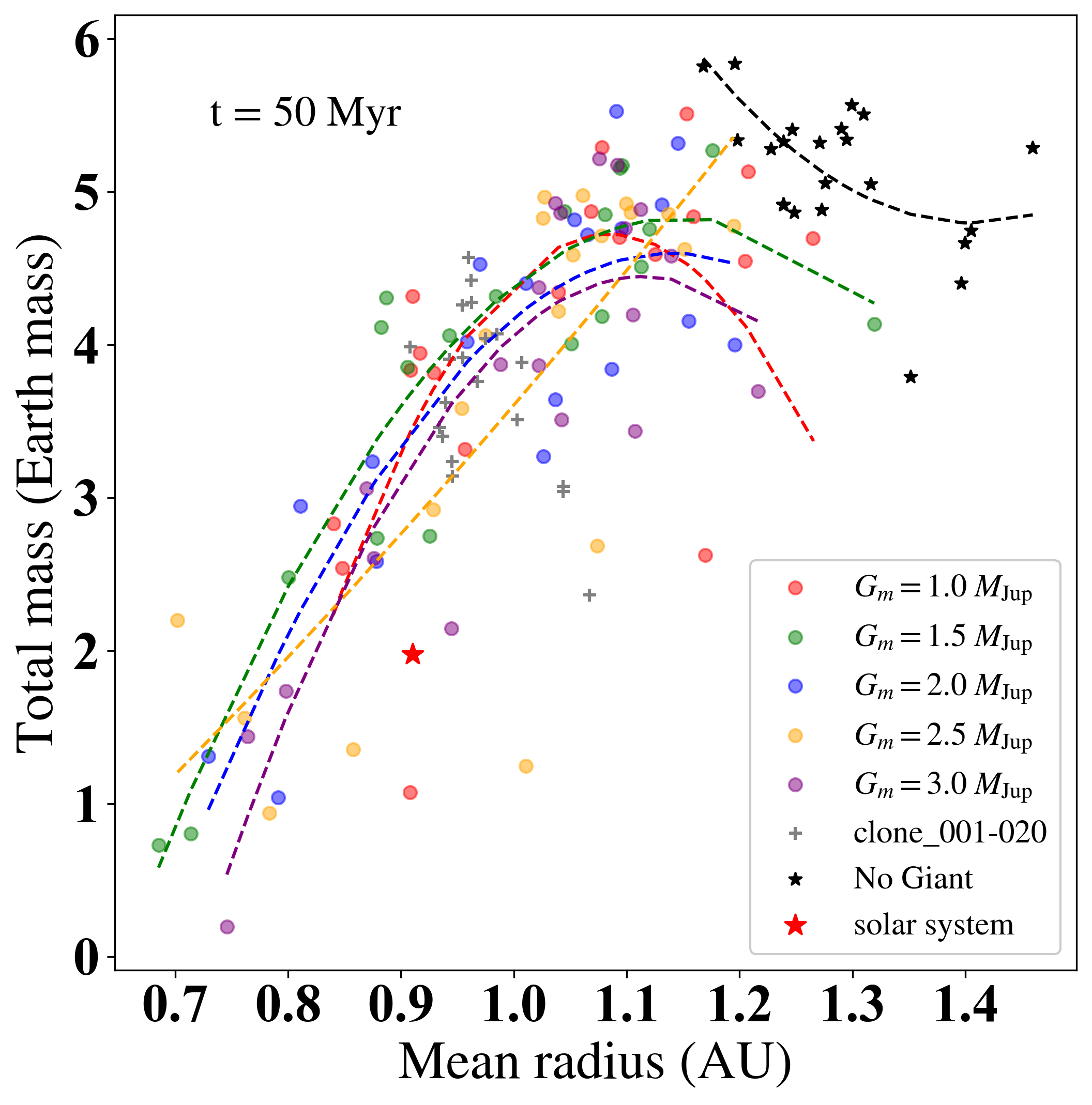}
        \caption{The plot shows the total mass of terrestrial planets as a function of the mean orbit for all simulation samples. The black stars represent the \textbf{No Giant} group and the red star represents the Solar System. The figure indicates that simulations setups containing a gas giant and those without gas giants fall into two distinct groups.}
        \label{pl_008}
    \end{figure}

Finally, Fig. \ref{pl_008} displays the total mass as a function of the mean orbit for all simulation samples. In this figure, we use black stars to represent the \textbf{No Giant} group and red star represents the Solar System. The figure shows that samples containing gas giants, \textbf{With Giant}, and samples \textbf{No Giant} are divided into two distinct groups. Due to the gravitational influence of the outer gas giant, the inner terrestrial planets exhibit relatively low residual mass and mean orbital radius. As the gas giant gradually moves away from the planetesimal disk, its gravitational impact weakens, leading to an increase in both mass and orbital width of these inner terrestrial planets. For the simulations \textbf{No Giant} located in the upper-right corner of the image, the terrestrial planets formed without the influence of an outer gas giant, resulting in more widely dispersed orbits and a greater residual mass.

\subsection{Gap Complexity}

A previous study by \cite{He_2023} used the catalog from the Kepler Giant Planet Search (KGPS) to study the relation between inner planets with the outer giant, and they use the $gap\, complexity$, $C$, to describe this orbital diversity \citep{Gilbert_2020}. The $gap\, complexity$ is defined as
\begin{center} 
\begin{equation}
\begin{aligned}
\mathcal{C} & \equiv-K\left(\sum_{i=1}^n p_i^* \log p_i^*\right) \cdot\left(\sum_{i=1}^n\left(p_i^*-\frac{1}{n}\right)^2\right)
\end{aligned}
\end{equation}
\end{center}
Here $n$ is the number of adjacent planet pairs (i.e. gaps) in the system, therefore at least three planets are required to compute the $C$. Also, $K$ is a multiplicity-dependent normalization constant chosen such that $C$ is always in the range (0, 1). The exact value of $K$ needs to be calculated according to \cite{Anteneodo_1996}. Here we use the approximate value, defined as follows
\begin{center} 
\begin{equation}
\begin{aligned}
K & \approx \frac{1}{0.262\ln(0.766n)}
\end{aligned}
\end{equation}
\end{center}
The quantity $p_i^*$ represents the orbital period information of planet, as defined below 
\begin{equation}
\begin{aligned}
p_i^* &= \frac{\log \mathcal{P}_i}{\log \left(\frac{P_{\max }}{P_{\min }}\right)}
\end{aligned}
\end{equation}

The quantity $\mathcal{P}_i \equiv P_{i+1} / P_i$ denotes the period ratios, while $P_{\min}$ and $P_{\max}$ are the minimum and maximum periods in the system. The gas complexity is designed to quantify the degree of departure from perfectly regular spacings  within a system, ranging from 0 (indicating evenly spaced planets in log-period) to 1 (representing maximum complexity).

    \begin{figure}
        \centering
        \includegraphics[width=0.5\textwidth]{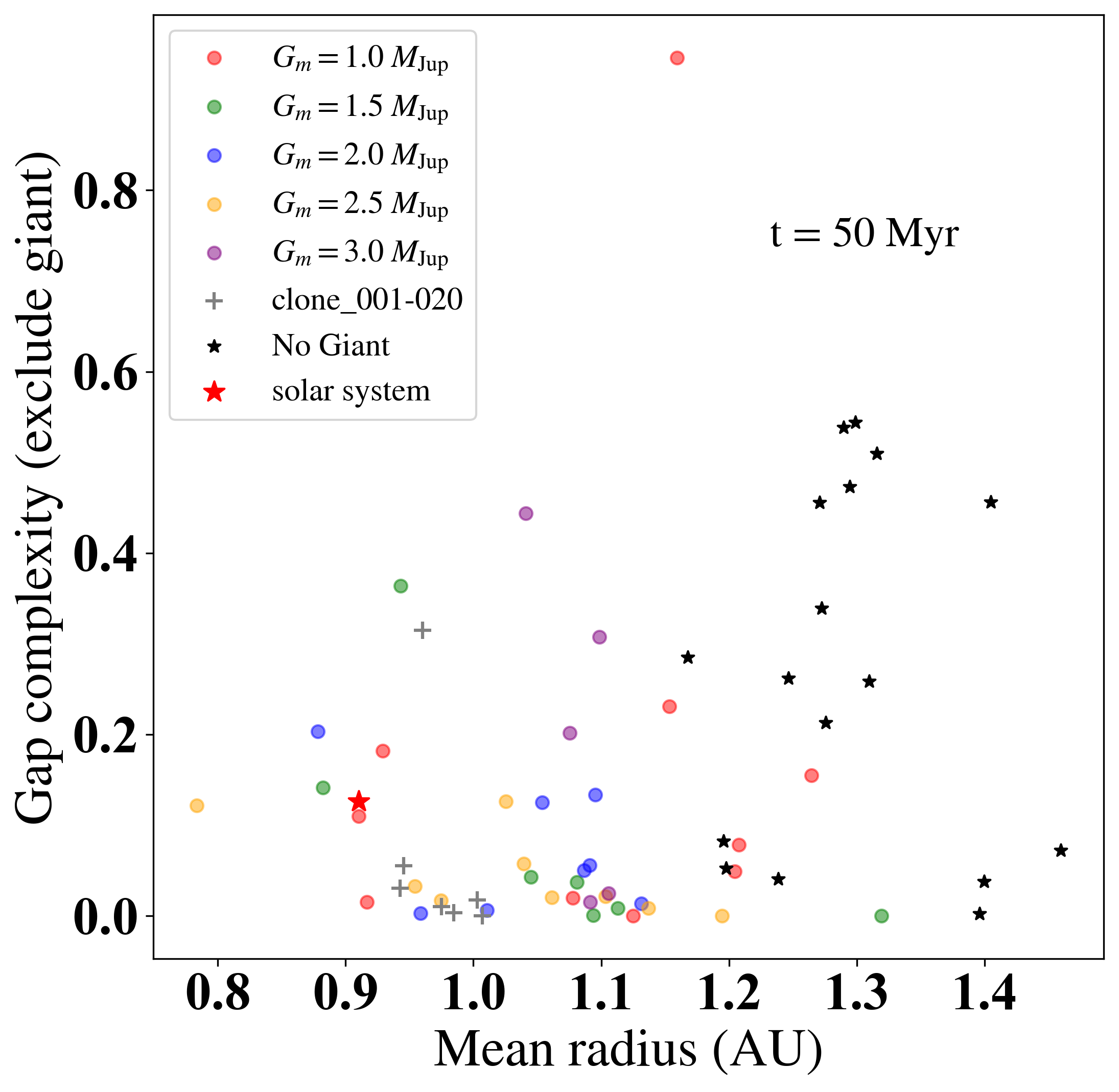}
        \caption{The figure shows the gap complexity, excluding the giant planet, as a function of the mean orbital radius of the inner planets. The black stars represent the \textbf{No Giant} group and the red star represents the Solar System. The figure indicates that the gap complexity within the inner multi-planet system is relatively low due to the influence of the outer giant's gravity.}
        \label{pl_009}
    \end{figure}

 \cite{He_2023} showed that observed systems with an outer giant planet have a higher gap complexity of the inner planets. However, the computation of gap complexity, $C$, necessitates the presence of at least three planets within the system. Excluding the remote giant planets, only 26 planetary systems from the Kepler Giant Planet Search (KGPS) satisfy this criterion, featuring at least three inner planets. These systems were subsequently categorized into 22 samples without an outer gas giant, 4 samples with on outer giant. However, the limited size of the observational sample presents challenges in discerning whether structural disparities in multi-planet systems arise from physical processes or incomplete observations. Therefore, we aim to employ simulations to investigate gap complexity as a quantitative measure of diversity within planetary systems. Firstly, we remove particles of less than $0.1\, M_\oplus$ and the outer giant when calculating the gap complexity, $C$. That minimizes the influence of low-mass particles and the outer giant in the results. We show the results in Fig. \ref{pl_009}. In the figure, the gap complexity excluding the giant planet is shown as a function of the mean orbital radius of the inner planets. The black stars represent the \textbf{No Giant} group and the red star represents the Solar System. However, due to the requirement to include at least 3 planets when calculating gap complexity, $C$, this leads to a reduction in the available samples. By analyzing these data we can find that the gap complexity within the inner multi-planet system is relatively low due to the influence of the outer giant's gravity.

    \begin{figure}
        \centering
        \includegraphics[width=0.5\textwidth]{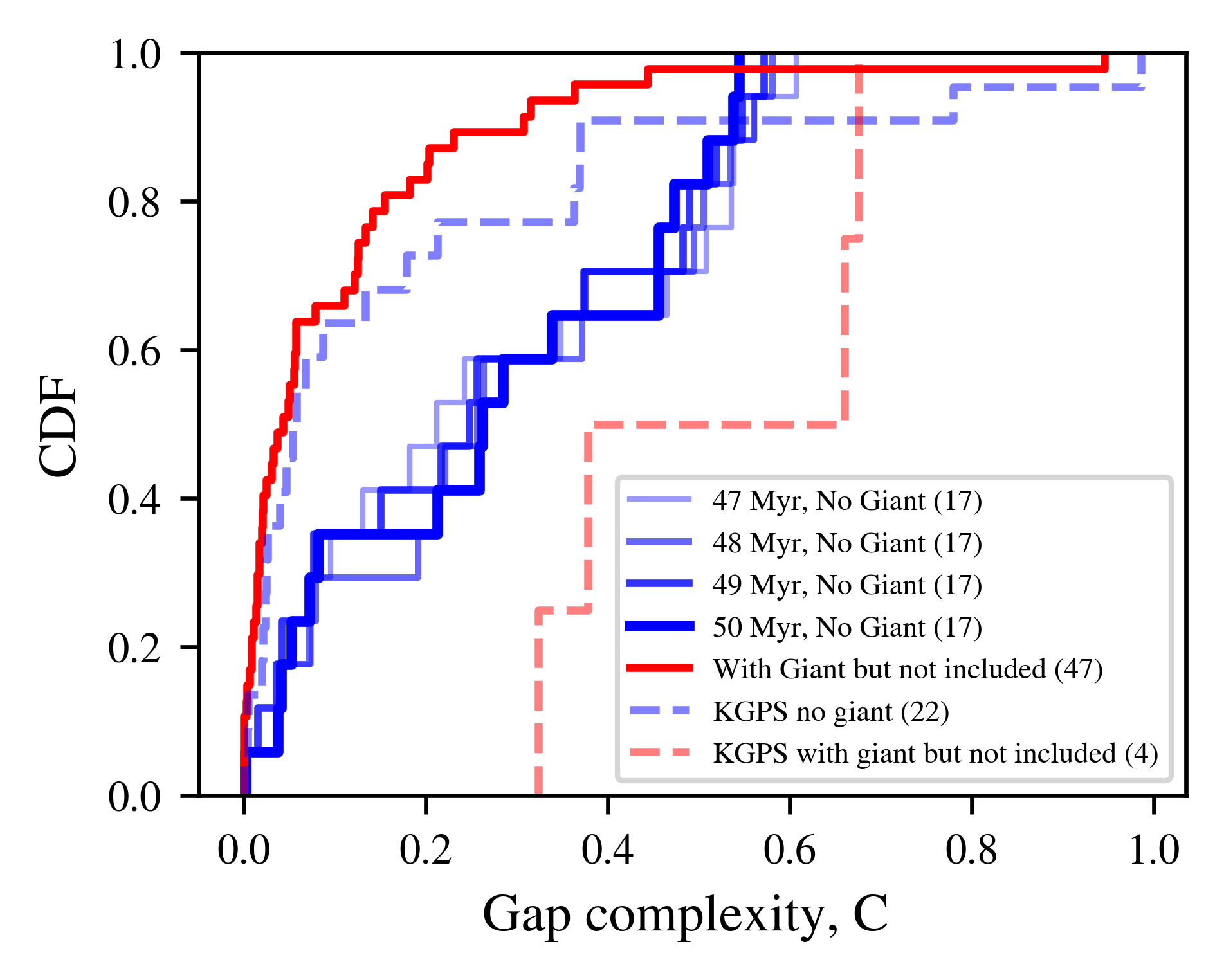}
        \caption{The figure shows the cumulative distributions of gap complexity. The blue lines represent \textbf{No Giant} group, while gradient and thickness colors are used to represent time from 47 to 50 Myr. The red line represent \textbf{With Giant} group. The blue and red dashed lines indicate the results without and with giants in the observed sample KGPS, respectively. All samples exclude giant from gap complexity calculations. The legend indicates the number of available samples in each sample set.}
        \label{pl_010}
    \end{figure}
    
Next, we compared the cumulative distributions of gap complexity in our simulated samples with the KGPS samples, as shown in Fig. \ref{pl_010}. The red line represents \textbf{With Giant} group at 50 Myr. For the \textbf{No Giant} group, the AMD (Angular Momentum Deficit) of the planetary system has not stabilized, as detailed in section 3.3, where we use gradient and thickness of blue lines representing results from 47 to 50 Myr. However, before calculating the gap complexity, we remove particles of less than $0.1\, M_\oplus$, therefore the evolution time has little effect on the cumulative distribution of the gap complexity. There are still significant differences between our simulated samples and the KGPS samples. For example, the trend of the \textbf{With Giant} samples is similar to the no giant samples in KGPS. Here, we present two possible causes to explain the difference between simulation and observation. a) Due to the limitations in the observing time of Kepler, outer gas giants in multi-planet systems cannot be directly observed. Consequently, the KGPS no giant samples (blue dashed line) could be mixed with outer giants that are not observed. As a result, the distribution function (CDF) curve of KGPS no giant (blue dashed line) falls between the simulated samples \textbf{With Giant} (red line) and \textbf{No Giant} (blue line). In the simulation samples, the gravity of the outer giant influences the orbit distribution of the inner multi-planet system, resulting in a relatively uniform distribution and low gap complexity values. Furthermore, the KGPS with giant samples (red dashed line) only contain four available samples, rendering this subset of results statistically insignificant. b) In the KGPS no giant samples, the super-Earths may form a resonant chain during the process of pebble accretion and migration \citep{Terquem_2007,Ogihara_2009,Cossou_2014,Izidoro_2017,Izidoro_2021b,Kajtazi_2023}, which results in a low value of the gap complexity (blue dashed line). Gravitational perturbations from the outer giants can disrupt the resonant orbits of the inner planets, leading to an increase in the gap complexity (red dashed line). However, our simulation sample adopts the classical collision formation model, which makes it unable to reproduce the results of the compact resonant chain formation pathway. To further elucidate the inconsistencies between the simulations and observations depicted in Fig. \ref{pl_010}, future terrestrial planet formation simulations should consider incorporating pebble accretion and migration during the protoplanetary disc phase.

\subsection{The statistical analysis}

    \begin{figure*}
        \centering
        \includegraphics[width=0.9\textwidth]{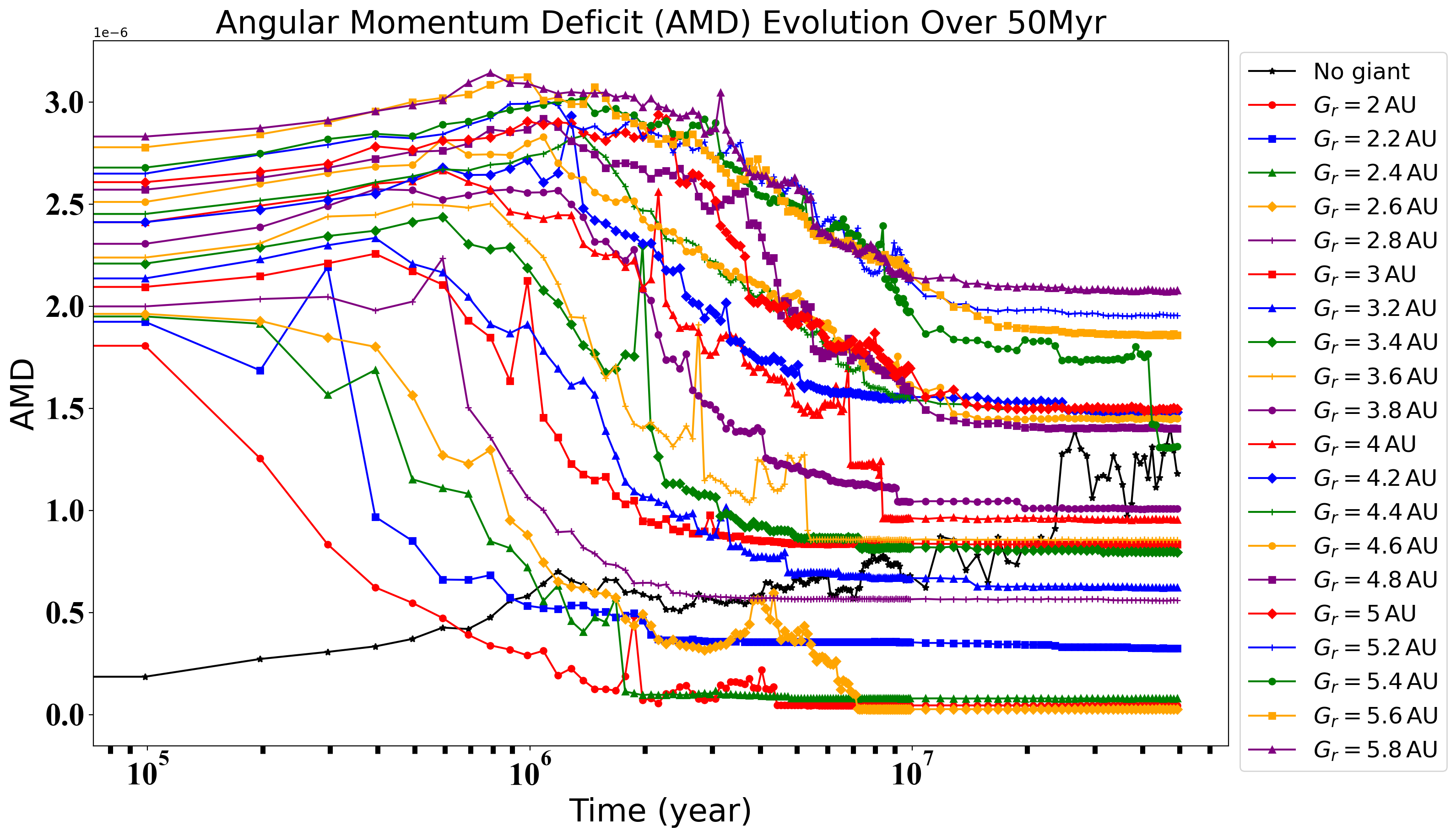}
        \caption{The figure shows the evolution of the angular momentum deficit (AMD) over 50 Myr in some samples (giant$001\text{--}020$ and nogiant\_000). The results indicate that the AMD represented by the black curve (\textbf{No Giant}) has not stabilized, whereas most of planetary systems have stabilized by 50 Myr.}
        \label{pl_011}
    \end{figure*}

\begin{figure}
   \centering
   \includegraphics[width=0.45\textwidth]{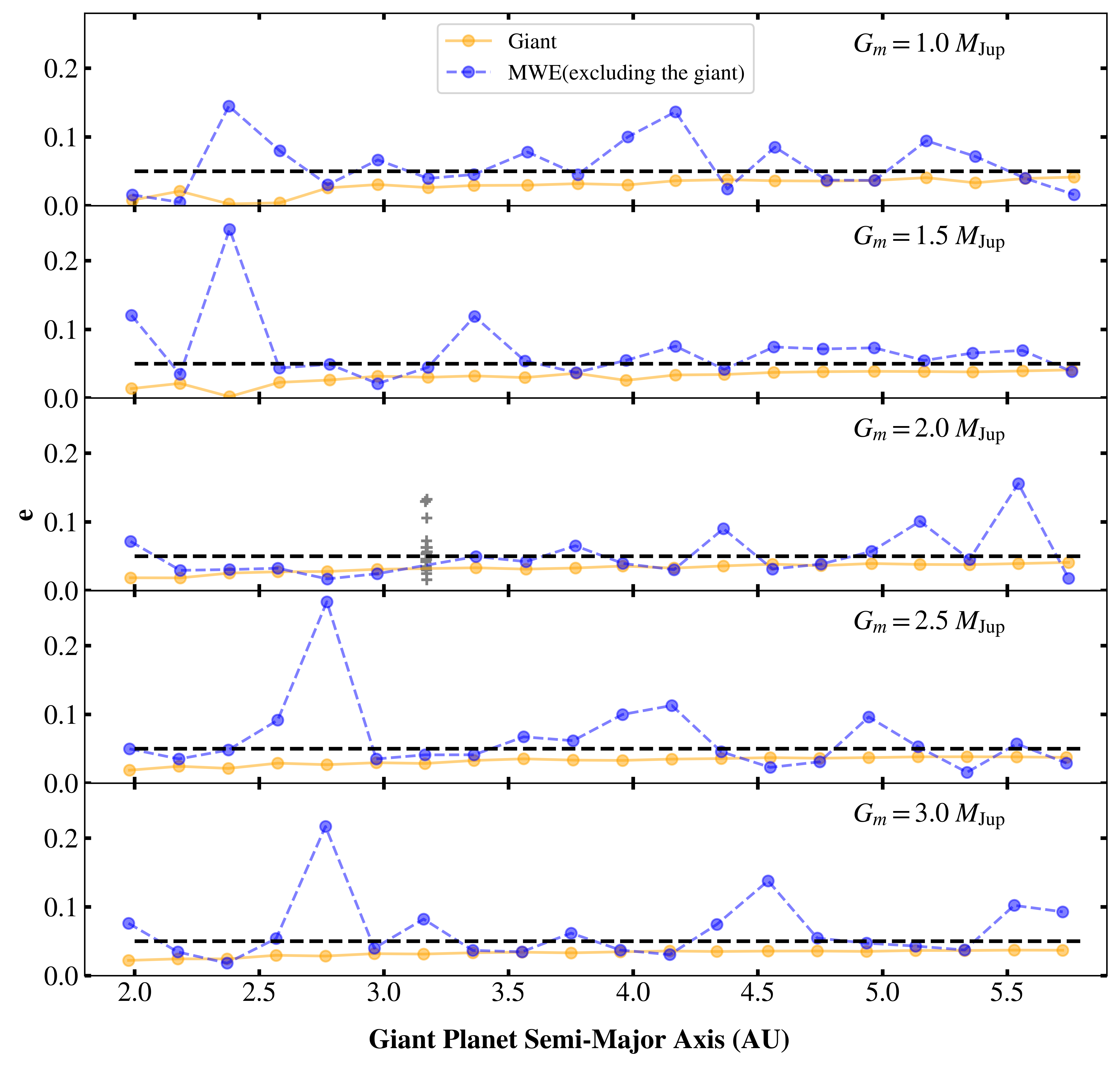}
   \caption{The figure shows the eccentricity of \textbf{With Giant} group at 50 Myr. The black dashed line represents the initial eccentricity of the giant, set at 0.05, while the orange dots depict the final giant eccentricity. The blue dots represent the mass weighted eccentricity, excluding the giant. The gray crosses represent the mass weighted eccentricity, excluding the giant, for the clone\_001-020.}
    \label{pl_012}
    \end{figure}

The concept Angular Momentum Deficit (AMD) plays a significant role in the formation of planetary systems \citep{Laskar_1997, Laskar_2000, Laskar_2017}. In studies of N-body dynamics simulations, collisions cease when the total AMD is insufficient to promote further planetary collisions. Therefore, we can use the total AMD to determine the evolution of planetary systems. In general, the total AMD is defined as follows:
\begin{equation}
\text{AMD} = \sum_{i} \lambda_i \left(1 - \sqrt{1 - e_i^2} \cdot \cos(i_i)\right)
\end{equation}
where the sum is over all planets in the system, $\lambda_i$ is the orbital angular momentum for the $i^{th}$ planet, $e_i$ represents the eccentricity of the $i^{th}$ planet and $i_i$ indicates the inclination of the $i^{th}$ planet's orbit relative to a reference plane. In Fig. \ref{pl_011}, we display the evolution of the angular momentum deficit (AMD) over 50 Myr in some samples (giant$001\text{--}020$ and nogiant\_000). The results indicate that the AMD represented by the black curve (\textbf{No Giant}) has not stabilized, whereas most of planetary systems have stabilized by 50 Myr. Even for those samples (\textbf{No Giant}) that are unstable, Fig. \ref{pl_010} shows little effect on the main results of this work. Here, we incorporate the outer gas giant into the AMD calculation. The eccentricity of this giant is influenced by the inner disk during the evolutionary process (seeing Fig. \ref{pl_002}), leading to a gradual increase in the AMD value with the distance of the giant from the inner disk. This phenomenon is explained in more detail in Fig \ref{pl_012}.

Next, we use the mass weighted eccentricity (MWE) from \cite{Raymond_2006} to calculate the eccentricities of the remaining particles at 50 Myr for all samples containing giants, as illustrated in Fig. \ref{pl_012}. The mass weighted eccentricity (MWE) is defined as follows:
\begin{equation}
    MWE = \frac{\sum_{j} m_j e_j}{\sum_{j} m_j},
\end{equation}
where we sum over all surviving bodies $j$ (excluding the giant).
In Fig. \ref{pl_012}, the black dashed line represents the initial eccentricity of the giant, set at 0.05, while the orange dots represent the final eccentricity of giant. The trend depicted by the orange curve indicates that as the orbital radius of the giant increases, its eccentricity is less affected by the inner disk. This results in a smaller deviation between the final eccentricity and the initial preset eccentricity. This trend, where the eccentricity of the giant planet increases with the increase of its orbital radius, is also reflected in Fig. \ref{pl_011} during the calculation of the Angular Momentum Deficit (AMD), which shows that the AMD increases as the giant planet's orbit becomes more distant. The blue dots represent the mass weighted eccentricity (MWE), excluding the giant. The peaks observed here signify that these samples possess a high eccentricity, and the influence of orbital resonance can be ruled out based on our orbital parameters. While the gray crosses suggest that the initial state of planetesimals and planetary embryos influences the final system's eccentricity, this deviation alone is insufficient to account for all the observed peaks. In fact, the snapshot depicting the evolution of the terrestrial planets presented in Fig. \ref{pl_002} corresponds to the first peak in the upper left corner of Fig. \ref{pl_012}. Consequently, the multiple terrestrial planets that ultimately form on the inner side are likely to increase the system's mass weighted eccentricity (MWE) due to gravitational interactions, leading to the observed peaks.

    \begin{figure}
        \centering
        \includegraphics[width=0.45\textwidth]{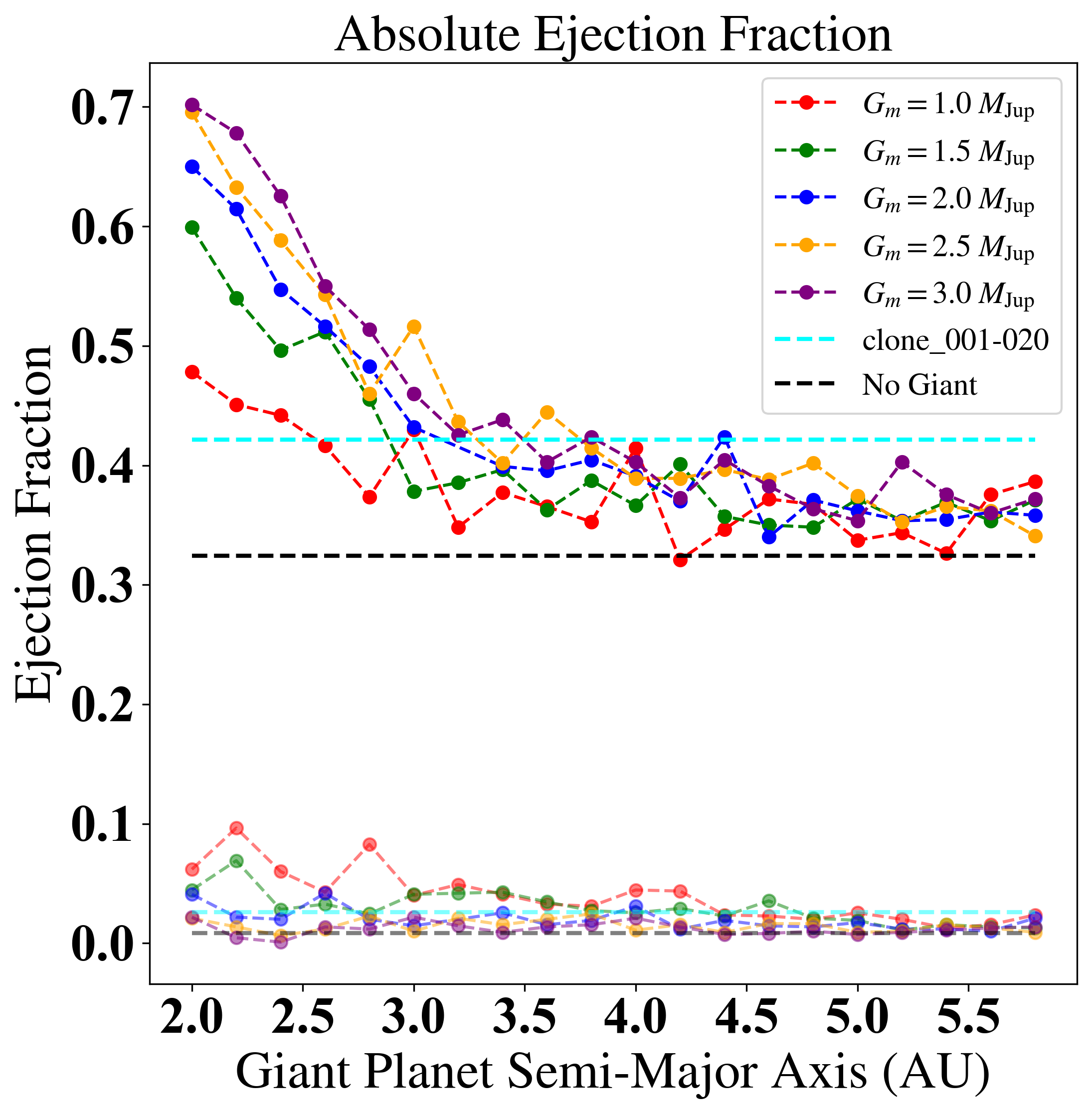}
        \caption{The figure shows the fraction of particles that are ejected from the system across all samples. The curves on the top represent the fraction of collisions with the central star (< 0.2 AU). The curves at the bottom represent the ejection fraction beyond the outer boundary (> 50 AU) of the system. The results indicate that the outer giant can significantly increase the particle ejection fraction.}
        \label{pl_013}
    \end{figure}
Subsequently, in Fig. \ref{pl_013} we show the fraction of particles that are ejected from the system across all samples. The curves at the top represent the fraction of collisions with the central star (< 0.2 AU). The curves at the bottom represent the ejection fraction beyond the outer boundary (> 50 AU) of the system. Comparing the black dashed line (\textbf{No Giant}) to the colored dashed lines (\textbf{With Giant}), the results indicate that the outer giant can significantly increase the particle ejection fraction. On the left side of Fig. \ref{pl_013}, the gravitational impact on the inner disk intensifies with the increasing mass of the giant, leading to a higher fraction of particles deviating from their original orbits and either colliding with the central star or being ejected from the system. Given that the gravitational force diminishes with the square of distance, the radius of the giant's orbit exerts a negligible influence on the ejection ratio of the inner particles beyond approximately 5 AU. In the \textbf{No Giant} group, approximately 22.9\% of the particles experienced planetesimal-to-planetesimal collisions before crashing into the central star. Therefore, in the model with fully active particles, collisions between planetesimals may increase the proportion of particles deviating from their original orbits and crashing into the central star (< 0.2 AU), even without the influence of the outer giant. In addition, upon comparing Fig. \ref{pl_012} and \ref{pl_013}, we find a strong correlation between the decrease in giant eccentricity and the mass of the ejected planetesimals and embryos from the system. As the giant orbits closer to the inner planetary disk, more mass is ejected from the system (< 0.2 AU or > 50 AU), resulting in a decrease in the eccentricity of the giant.
    \begin{figure}
        \centering
        \includegraphics[width=0.45\textwidth]{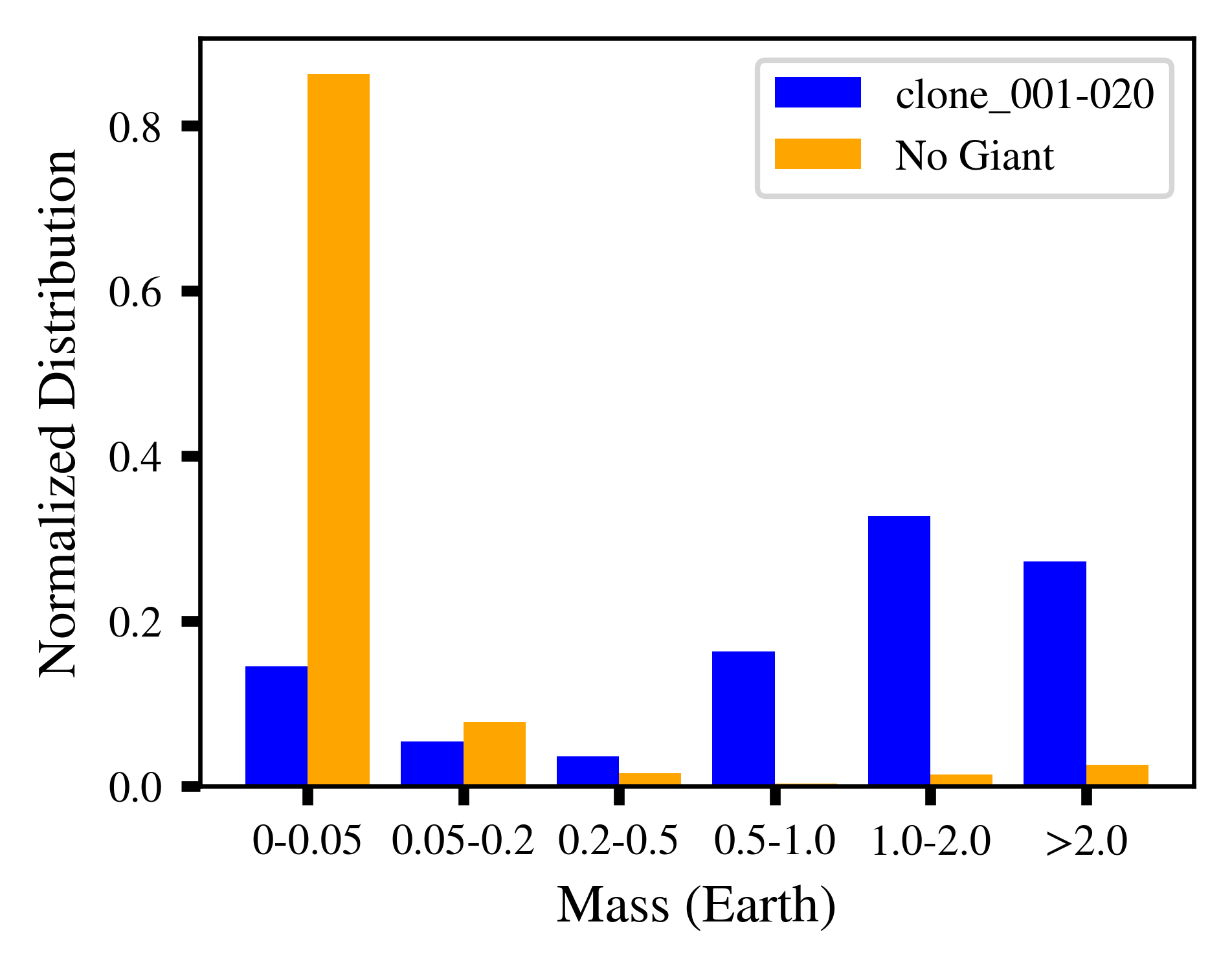}
        \caption{The figure shows a statistical histogram depicting the mass distribution of terrestrial planets and remaining planetesimals. The blue bars represent the clone\_001-020, which includes a giant planet fixed at 3.2 AU, while the orange bars represent the \textbf{No Giant} group. This suggests that the gravitational influence of an outer gas giant actually facilitates the formation of massive terrestrial planets by affecting the interactions between planetesimals and planetary embryos in the terrestrial planet zone.}
        \label{pl_014}
    \end{figure}

\begin{figure}
        \centering
        \includegraphics[width=0.45\textwidth]{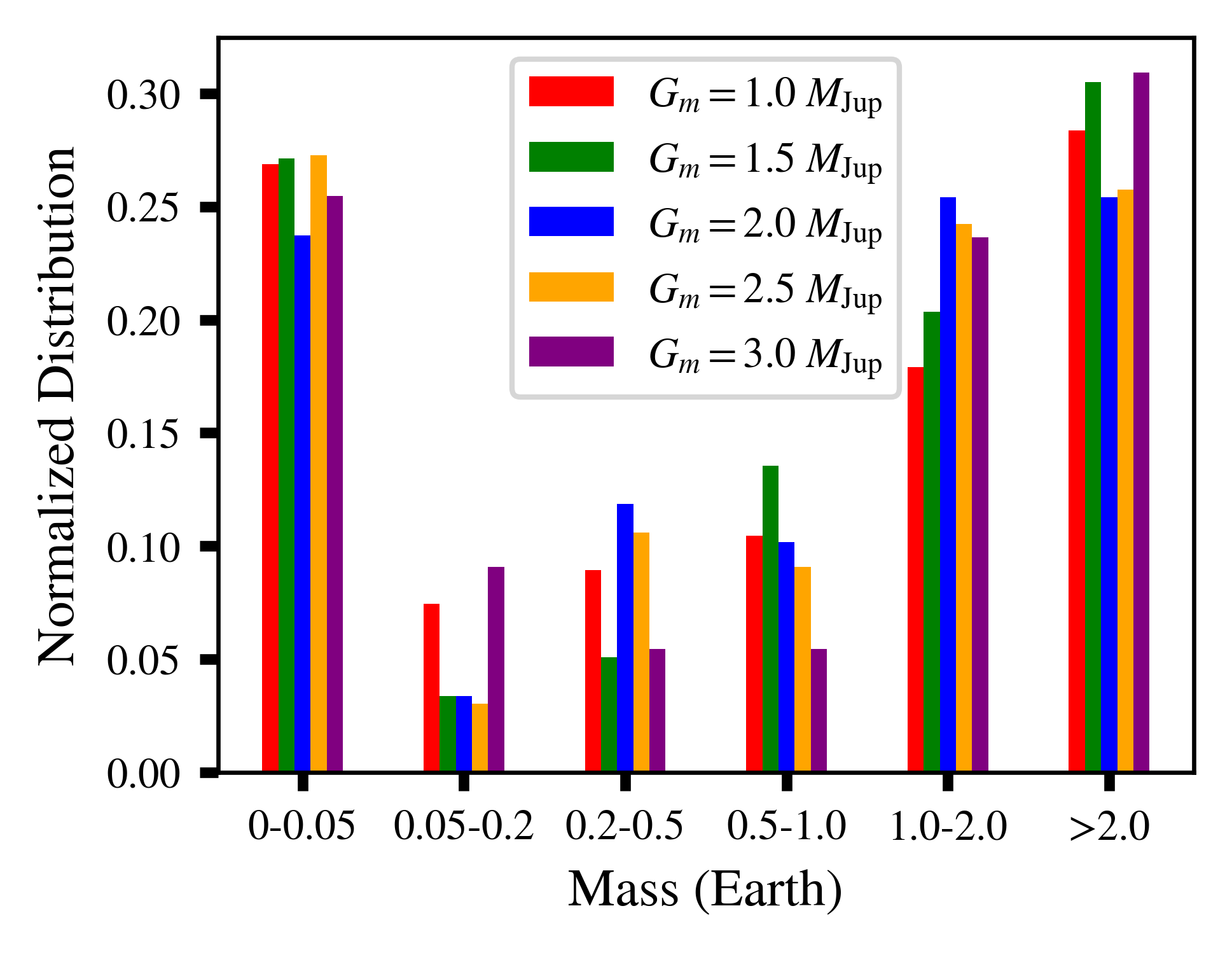}
        \caption{The figure shows a statistical histogram depicting the mass distribution of terrestrial planets and the remaining planetesimals. We have categorized the samples into five groups based on the initial mass of the giants, each represented by a distinct colors. Comparing with the \textbf{No Giant} in Fig. \ref{pl_014}, the result shows that the presence of more massive gas giants in the outer region of the planetesimal disk significantly increase the likelihood of forming massive terrestrial planets in the inner regions.}
        \label{pl_015}
    \end{figure}

Next, we show a statistical histogram of different masses of the terrestrial planets and remnant planetesimals in Fig. \ref{pl_014} and Fig. \ref{pl_015}. The particles that have not gained significant mass (initial planetesimals and embryos) are indicated by column $0\text{--}0.05\, M_\oplus$. In Fig. \ref{pl_014}, the blue bars represent the samples of the clone\_001-020, $G_{\text{m}}=2.0 \, M_{\rm Jup}$ and $G_{\text{a}}=3.2$\,AU and the orange bars represent the samples of \textbf{No Giant}. The findings reveal that when there is no gas giant in an exterior orbit (orange data), most particles fail to undergo mass growth through mutual collisions, and even a minimal increase in mass does not lead to the formation of a massive planet. However, the presence of a an exterior gas giant (blue data) leads to a higher number of particles involved in the collision process, resulting in the formation of massive planets. This suggests that the gravitational influence of the outer gas giants facilitates the formation of massive terrestrial planets through their impact on the interaction between planetesimals and planetary embryos within planetesimals.

In Fig. \ref{pl_015}, we shows a statistical histogram of different mass sizes of the terrestrial planets and remnant planetesimals on \textbf{With Giant}, and we divided the samples into five categories based on the mass of the giant planet, represented by different colors. However, there are no significant differences in the distribution patterns of these five categories. With the increase in the mass of the giant, there is only a slight decrease in the proportion of planets in the 0.5-1.0 $M_\oplus$ range and a slight increase in the proportion of planets in the 1.0-2.0 $M_\oplus$ range. The figure shows that when the planetary system evolves to a time of 50 Myr, only about 25\% of planetesimals remain with a mass below $0.05\, M_\oplus$, which means that the mass of these particles have not increased significantly. Comparing with the \textbf{No Giant} group in Fig. \ref{pl_014}, the result shows that the presence of gas giants in the outer region of the planetesimal disk enhances the likelihood of forming massive terrestrial planets on the inside.

    \begin{figure}
        \centering
        \includegraphics[width=0.45\textwidth]{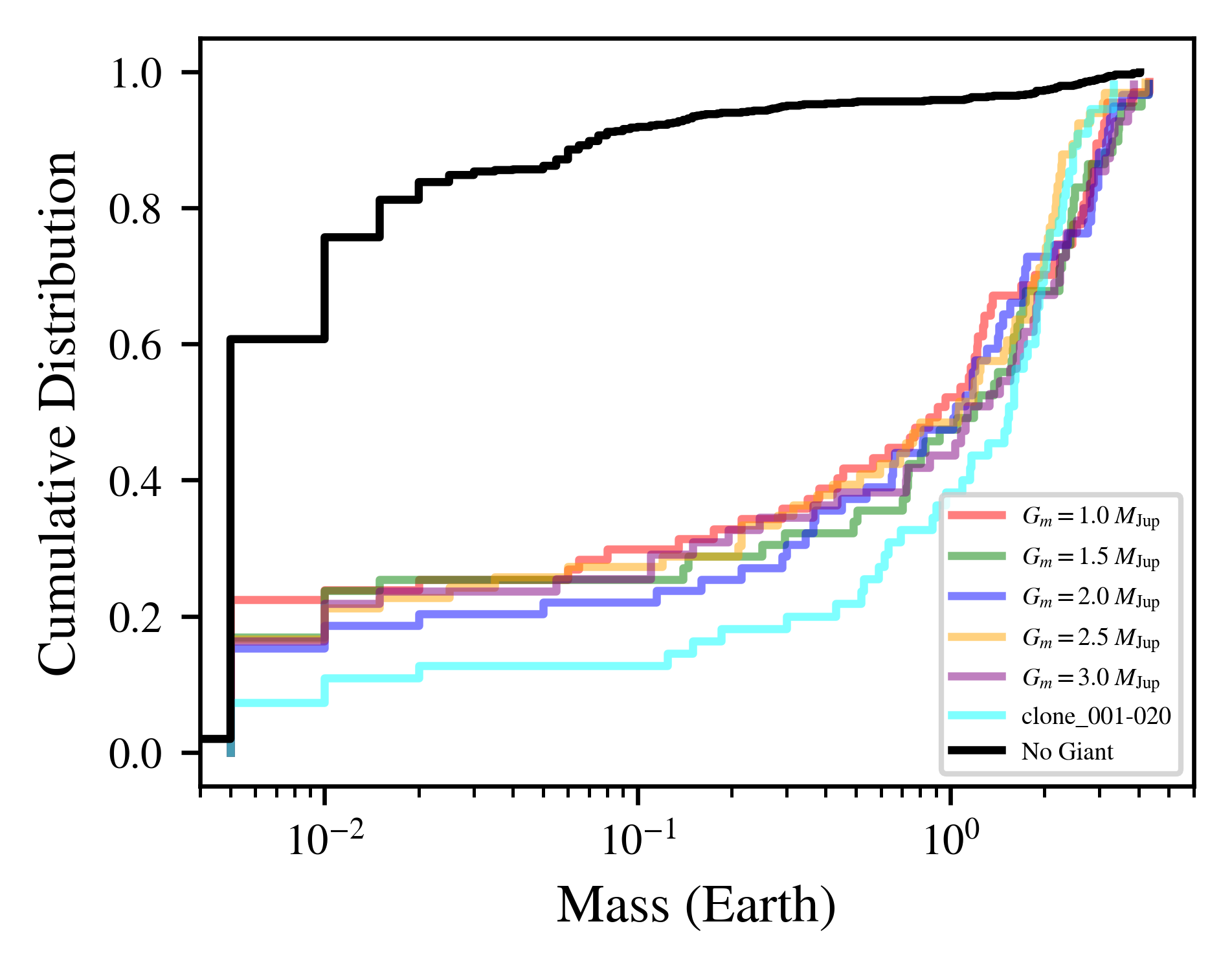}
        \caption{The figure shows a cumulative distribution of both minor and major planetary bodies that formed in our simulations, with different colors representing distinct sample sets.}
        \label{pl_016}
    \end{figure}

Finally, we show a cumulative distribution of both minor and major planetary bodies that formed, using all the samples we studied, in Fig. \ref{pl_016}. We use different colors representing different sample sets. The curves in the figure are categorized into three groups based on different sample sets. The black curve at the top represents samples of \textbf{No Giant}, the bottom cyan curve represents samples of the  different initial states of the planetesimals and planetary embryos clone\_001-020, and the middle samples, giant\_001-100, are divided into five different curves, depending on the mass of the initial gas giant. The results here are consistent with Fig. \ref{pl_014} and \ref{pl_015} showing that gas giants on the outside of the planetesimals disk facilitates the formation of massive terrestrial planets on the inside.

\section{Conclusions}

We end here with a brief conclusion in section 4.1, where we present a comprehensive summary of the simulation results for two distinct types of samples. We investigated the influence of outer gas giants on the formation of a system of terrestrial planets in giant\_001-100; we took the same initial state of a gas giant as giant\_047 and randomly generate planetesimals and embryos in clone\_001-020; we represented scenarios where only planetesimals and embryos are present, without gas giants, in nogiant. In section 4.2, we succinctly outline the limitations of our model and provide a forward-looking perspective on potential avenues for future research endeavors.

\subsection{Summary of our simulation results}

   \begin{enumerate}
      \item \textbf{With Giant}: Fig. \ref{pl_003} shows the result of planetary system simulation of approximately 50 Myr. By this stage, only a few massive particles remain, forming the final architecture of the planetary system \citep{Levison_2003, Raymond_2004, Raymond_2006}. By comparing nogiant\_000 with giant\_001-100, it becomes evident that the presence of an outer giant leads to more compact orbital configuration for the inner terrestrial planets. Meanwhile, a gas giant too close to the planetesimal disk entirely prevents the formation of terrestrial planets. Also, the number and size of the remaining planets in planetary systems are relatively limited for a close-in giant planet. Moving the gas giant further away from the planetesimal disk will increase the diversity of terrestrial planetary. Meanwhile, we show the evolution of the angular momentum deficit (AMD) over 50 Myr is some samples (giant$001\text{--}020$) in Fig. \ref{pl_011}. The results indicate that the AMD represented the \textbf{No Giant} sample has not stabilized, whereas most of planetary systems have stabilized by 50 Myr. While keeping the conditions of the outer gas giant constant (clone\_001-020), the findings demonstrate that random perturbations to the initial state of planetesimals and planetary embryos exert significant influences on the orbits and masses of the resultant terrestrial planets. However, the initial conditions of planetesimals and planetary embryos seems to have little effect on their final total mass and mean orbit of the inner terrestrial planets in Fig. \ref{pl_004}. Therefore, the total mass and mean orbital characteristics of the inner remnant terrestrial planets in a planetary system can serve as robust physical parameters for investigating the formation process of such planets.\\ \\
      \item \textbf{No Giant}: For models that do not harbour an outer gas giant, our findings are consistent with the fixed-giant samples, clone\_001-020, where small changes in the initial state of planetesimals and planetary embryos have a significant impact on the results in Fig. \ref{pl_005}. Furthermore, this effect can be significantly reduced by calculating the total mass and mean orbit of the remaining planet. \\ \\
      \item The effect of gas giants: Under the gravitational influence of a outer giants, collisions between inner planetesimals and embryos are promoted, which accelerates the formation process of terrestrial planets in Fig. \ref{pl_011} while also ejecting a large number of particles out of the planetary system in Fig. \ref{pl_013}. Therefore, when the giant is close to the inner disk, the total mass and mean orbital radius of the ultimately formed terrestrial planets are smaller. When the giant is moved away from the inner disk to a distance of 5 AU, due to the weakened gravitational influence of the giants, the total mass and mean orbital radius of the resulting terrestrial planets becomes larger in Fig. \ref{pl_003}. Moreover, the orbit of the giant itself is modulated by the gravitational pull from the inner disk, resulting in a gradual reduction in the eccentricity of the giant's orbit in Fig. \ref{pl_002}. However, this modulatory effect diminishes as the orbital radius of the giant extends away from the inner disk in Fig. \ref{pl_012} \citep{Levison_2003,Raymond_2006}. In addition, upon comparing Fig. \ref{pl_012} and \ref{pl_013}, we find a strong correlation between the decrease in giant eccentricity and the mass of the ejected planetesimals and embryos from the system. As the giant orbits closer to the inner planetary disk, more mass is ejected from the system (< 0.2 AU or > 50 AU), resulting in a decrease in the eccentricity of the giant.\\ \\
      \item Gap complexity: Under the classical theory, where planetary embryos and planetesimals collide to form terrestrial planets, our simulation results do not reach the same conclusion as \cite{He_2023} that the presence of an outer giant increases the gap complexity of the planetary system. In contrast, our results show that the presence of a giant planet actually decreases the gap complexity. We believe that this difference between these two results may arise due to different planet formation scenarios. The systems of super-Earth and mini-Neptunes analysed by \cite{He_2023} likely formed by within the protoplanetary disc and experienced significant migration and subsequent planetary instabilities \citep{Lambrechts_2019,Izidoro_2021b,Kajtazi_2023}. The classical terrestrial planet formation simulations employed here, on the other hand, yield smaller planets that are not subject to disc migration, due to their formation after the dissipation of the protoplanetary disc. \\ \\
      \item The statistical analysis: The implication of this analysis is that when the gas giant is close to the planetesimal disk, its gravitational influence is amplified and this expedites the collision or scattering of planetesimals and planetary embryos, resulting in a rapid reduction in the number of particles in the system. When the gas giant is moved away from the planetesimal disk, its gravitational influence weakens, resulting in reduced collisions between planetesimals and planetary embryos. Moreover, Fig. \ref{pl_014}, \ref{pl_015} and \ref{pl_016} show that a gas giant on the outside of the planetesimal disk facilitates the formation of massive terrestrial planets on the inside. However, if the giants are too close to the disk of planetesimals and embryos, this would prevent the formation of large-sized terrestrial planets, see Fig. \ref{pl_003}. 
   \end{enumerate}

\subsection{Future research}

In our study, the cost of our N-body simulation model imposes stringent constraints on the possibility to vary our initial parameters of inner planetesimals and planetary embryos. Specifically, in giant\_001-100, 
we chose to keep the initial conditions for planetesimals and planetary embryos identical. As a result, our conclusions are subject to the limitation that we did not explore a wider range of initial conditions for planetesimals and planetary embryos. In nogiant\_000-020 and clone\_001-020, we nevertheless briefly explore the impact of randomized variations in the initial positions of inner planetesimals and planetary embryos on the simulation results. These changes were seen to have an impact on the specific outcome of the simulations, but not on  averaged properties such as mean mass and position. In addition, the narrow model we use, which concentrates planetesimals and planetary embryos around 1 AU in the initial conditions, makes it difficult for us to discuss the sources of water in planets and the relationship between different gas giant orbits and planet formation in the habitable zone, compared to the wide condition \citep{Raymond_2006c}. Furthermore, migration of the gas giant was omitted in our study. Previous research has demonstrated that the migration and instability of gas giants can significantly influence the dynamics of inner planets \citep{Fogg_2005,Fogg_2007,Raymond_2006b,Edward_2008, Nesvorn_2021}. We also ignored the growth of the embryos by pebble accretion. Thus, incorporating other accretion mechanisms and various gas giant migration models will be a focal point in our future investigations.

\begin{acknowledgements}
    We would like to thank the referee, Sean Raymond, for many comments that helped improve the manuscript. Z.K. thanks the support by the National Natural Science Foundation of China under Grants Nos. 11920101003, 12021003, and 11633001, and the Strategic Priority Research Program of Chinese Academy of Sciences, Grant No.XDB23000000. A.J. acknowledges funding from the Danish National Research Foundation (DNRF Chair Grant DNRF159), the Carlsberg Foundation (Semper Ardens: Advance grant FIRSTATMO), the Knut and Alice Wallenberg Foundation (Wallenberg Scholar Grant 2019.0442) and the Göran Gustafsson Foundation. M.L. acknowledges funding from the European Research Council (ERC Starting Grant 101041466-EXODOSS). The Tycho supercomputer hosted at the SCIENCE HPC center at the University of Copenhagen was used for supporting this work.
\end{acknowledgements}

%
%
\bibliographystyle{aa}
\bibliography{mybibliography}

\end{document}